\documentclass{achemso}
\pdfoutput=1
\usepackage[version=3]{mhchem} 
\usepackage[T1]{fontenc} 
\usepackage[english]{babel}
\usepackage{mathtools}
\usepackage{xfrac}
\usepackage{graphicx}
\usepackage[separate-uncertainty = true]{siunitx}
\usepackage{hyperref}
\usepackage{subcaption}
\usepackage[dvipsnames]{xcolor}
\usepackage{soul}
\usepackage{upgreek}
\usepackage{caption}
\usepackage{lineno}

\newcommand{\ML}{1L-MoS\textsubscript{2}}
\author{Tobias Bucher}
\affiliation{Institute of Solid State Physics, Friedrich Schiller University Jena, 07743 Jena, Germany}
\alsoaffiliation{Institute of Applied Physics, Friedrich Schiller University Jena, 07745 Jena, Germany}
\alsoaffiliation{Abbe Center of Photonics, Friedrich Schiller University Jena, 07745 Jena, Germany}
\altaffiliation{These authors contributed equally}
\email{tobias.bucher@uni-jena.de}
\author{Zlata Fedorova}
\affiliation{Institute of Solid State Physics, Friedrich Schiller University Jena, 07743 Jena, Germany}
\alsoaffiliation{Institute of Applied Physics, Friedrich Schiller University Jena, 07745 Jena, Germany}
\alsoaffiliation{Abbe Center of Photonics, Friedrich Schiller University Jena, 07745 Jena, Germany}
\altaffiliation{These authors contributed equally}
\email{zlata.fedorova@uni-jena.de}
\author{Mostafa Abasifard}
\affiliation{Institute of Applied Physics, Friedrich Schiller University Jena, 07745 Jena, Germany}
\alsoaffiliation{Institute of Solid State Physics, Friedrich Schiller University Jena, 07743 Jena, Germany}
\alsoaffiliation{Abbe Center of Photonics, Friedrich Schiller University Jena, 07745 Jena, Germany}
\author{Rajeshkumar Mupparapu}
\affiliation{Institute of Applied Physics, Friedrich Schiller University Jena, 07745 Jena, Germany}
\alsoaffiliation{Abbe Center of Photonics, Friedrich Schiller University Jena, 07745 Jena, Germany}
\author{Matthias J. Wurdack}
\affiliation{ARC Centre of Excellence in Future Low-Energy Electronics Technologies and Department of Quantum Science and Technology, Research School of Physics, The Australian National University, Canberra, ACT, 2601, Australia}
\alsoaffiliation{Institute of Solid State Physics, Friedrich Schiller University Jena, 07743 Jena, Germany}
\alsoaffiliation{Institute of Applied Physics, Friedrich Schiller University Jena, 07745 Jena, Germany}
\alsoaffiliation{Abbe Center of Photonics, Friedrich Schiller University Jena, 07745 Jena, Germany}
\author{Emad Najafidehaghani}
\affiliation{Institute of Physical Chemistry, Friedrich Schiller University Jena, 07743 Jena, Germany}
\author{Ziyang Gan}
\affiliation{Institute of Physical Chemistry, Friedrich Schiller University Jena, 07743 Jena, Germany}
\author{Heiko Knopf}
\affiliation{Fraunhofer Institute for Applied Optics and Precision Engineering IOF, 07745 Jena, Germany}
\alsoaffiliation{Institute of Applied Physics, Friedrich Schiller University Jena, 07745 Jena, Germany}
\alsoaffiliation{Abbe Center of Photonics, Friedrich Schiller University Jena, 07745 Jena, Germany}
\alsoaffiliation{Max Planck School of Photonics, Germany}
\author{Antony George}
\affiliation{Institute of Physical Chemistry, Friedrich Schiller University Jena, 07743 Jena, Germany}
\alsoaffiliation{Abbe Center of Photonics, Friedrich Schiller University Jena, 07745 Jena, Germany}
\author{Falk Eilenberger}
\affiliation{Fraunhofer Institute for Applied Optics and Precision Engineering IOF, 07745 Jena, Germany}
\alsoaffiliation{Institute of Applied Physics, Friedrich Schiller University Jena, 07745 Jena, Germany}
\alsoaffiliation{Abbe Center of Photonics, Friedrich Schiller University Jena, 07745 Jena, Germany}
\alsoaffiliation{Max Planck School of Photonics, Germany}
\author{Thomas Pertsch}
\affiliation{Institute of Applied Physics, Friedrich Schiller University Jena, 07745 Jena, Germany}
\alsoaffiliation{Abbe Center of Photonics, Friedrich Schiller University Jena, 07745 Jena, Germany}
\alsoaffiliation{Fraunhofer Institute for Applied Optics and Precision Engineering IOF, 07745 Jena, Germany}
\alsoaffiliation{Max Planck School of Photonics, Germany}
\author{Andrey Turchanin}
\affiliation{Institute of Physical Chemistry, Friedrich Schiller University Jena, 07743 Jena, Germany}
\alsoaffiliation{Abbe Center of Photonics, Friedrich Schiller University Jena, 07745 Jena, Germany}
\alsoaffiliation{Jena Center for Soft Matter (JCSM), 07743 Jena, Germany}
\author{Isabelle Staude}
\affiliation{Institute of Solid State Physics, Friedrich Schiller University Jena, 07743 Jena, Germany}
\alsoaffiliation{Institute of Applied Physics, Friedrich Schiller University Jena, 07745 Jena, Germany}
\alsoaffiliation{Abbe Center of Photonics, Friedrich Schiller University Jena, 07745 Jena, Germany}
\alsoaffiliation{Max Planck School of Photonics, Germany}
\title{Influence of resonant plasmonic nanoparticles on optically accessing the valley degree of freedom in 2D semiconductors}


\begin{document}
%

%
\newpage
\begin{abstract}
    The valley degree of freedom is one of the most intriguing properties of atomically thin transition metal dichalcogenides. Together with the possibility to address this degree of freedom by valley-contrasting optical selection rules, it has the potential to enable a completely new class of future electronic and optoelectronic devices. Resonant optical nanostructures emerge as promising tools for interacting with and controlling the valley degree of freedom at the nanoscale. However, a critical understanding gap remains in how nanostructures and their nearfields affect the circular polarization properties of valley-selective emission hindering further developments in this field. In order to address this issue, our study delves into the experimental investigation of a hybrid model system where valley-specific emission from a monolayer of molybdenum disulfide is interacting with a resonant plasmonic nanosphere. Contrary to the simple intuition suggesting that a centrosymmetric nanoresonator preserves the degree of circular polarization in the forward scattered farfield by angular momentum conservation, our cryogenic photoluminescence microscopy reveals that the light emitted from the nanoparticle position is largely unpolarized, i.e. we observe depolarization. We rigorously study the nature of this phenomenon numerically considering the monolayer-nanoparticle interaction at different levels including excitation and emission. In doing so, we find that the farfield degree of polarization strongly reduces in the hybrid system when including excitons emitting from outside of the system's symmetry point, which in combination with depolarisation at the excitation level causes the observed effect. Our results highlight the importance of considering spatially distributed emitters for precise predictions of polarization responses in these hybrid systems. This finding advances our fundamental knowledge of the light-valley interactions at the nanoscale but also unveils a serious impediment of the practical fabrication of resonant valleytronic nanostructures.
\end{abstract}
%
    %
\section{Introduction}
\begin{figure}[h]
        \centering
        \includegraphics[width=8cm]{"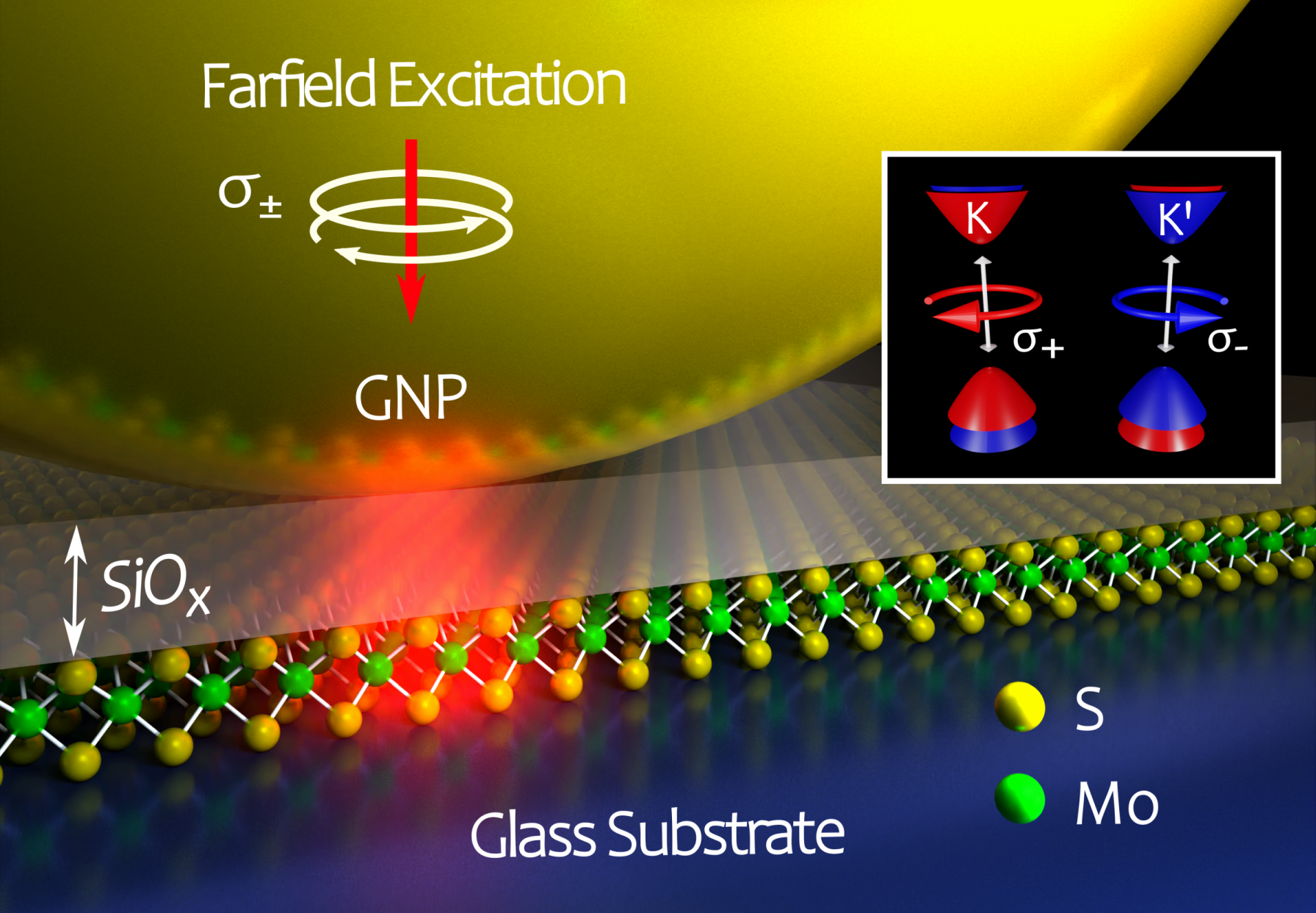"}
        \caption{\textbf{Schematic of the investigated hybrid system excited by circularly polarized light.} A resonant gold nanoparticle (size 200 nm) is placed above a monolayer of molybdenum disulfide situated on a glass substrate. A thin dielectric spacer layer of silicon oxide (thickness 15 nm) was introduced prior to the nanoparticle deposition in order to prevent direct metal-TMD contact. Note that the crystal structure was scaled up for better visibility. The inset illustrates the valley-contrasting optical selection rules in monolayer TMDs.}\label{fig:concept}
    \end{figure}

    As modern CMOS-based information technology is facing fundamental limits of further downscaling, novel materials providing additional electronic degrees of freedom, such as spin or valley-pseudospin, have become an active field of research~\cite{xu2014spin, sun2019separation, gong2018nanoscale, zhang2014electrically, hu2019coherent, jones2013optical}. The valley-pseudospin arises from multiple degenerate but inequivalent energy extrema in the bands of a crystal, in which excitons with distinct spin states can form that may be used to encode and process information~\cite{xiao2007valley, schaibley2016valleytronics, xiao2012coupled}. In two-dimensional transition metal dichalcogenides (2D-TMDs), the broken inversion symmetry of the crystal structure and strong spin-orbit coupling lead to spin-valley locking. Consequently, energy-degenerate excitons with opposite spin are located at the K/K$'$ points (or valleys) at the corners of the Brillouin zone~\cite{zeng2012valley, mak2012control,cao2012valley} following valley-contrasting optical selection rules. The pronounced direct bandgap photoluminescence (PL) in the monolayer phase~\cite{Splendiani2010emerging, Mak2010atomically} facilitates efficient addressing and readout of the valley-degree of freedom in 2D-TMDs using circularly polarized light as depicted by the inset in Fig.~\ref{fig:concept}. Photons with circular polarization $\sigma^{+}$ ($\sigma^{-}$) only interact with carriers in the valley K (K$'$) and the valley-selective occupation can be quantified by the degree of circular polarization, $\text{DOCP}=(\mathcal{I}_{\sigma^+}-\mathcal{I}_{\sigma^-})/(\mathcal{I}_{\sigma^+}+\mathcal{I}_{\sigma^-})$, based on the valley-selective PL intensities $\mathcal{I}_{\sigma^{\pm}}$. However, despite an efficient control knob for the valley-pseudospin by means of circularly polarized light, the robust detection, manipulation and transport of the valley-pseudospin information remains challenging, mainly because of the short lifetime of valley-polarized excitonic states and a strongly reduced DOCP at room temperature due to phonon-assisted intervalley scattering.\\
    During the last few years, the integration of 2D-TMDs with photonic nanostructures has gained immense popularity as an approach to address these challenges by enhancing and tailoring light-valley interaction at the nanoscale. In the field of chiral plasmonics, the modulation of the valley-pseudospin is commonly discussed on the basis of chiral Purcell enhancement~\cite{Yoo2014chiral} involving superchiral nearfields.~\cite{Tang2010optical} While such chiral metamaterials have been proven to selectively modulate the valley dynamics and can lead to a sizable DOCP from 2D-TMDs even at room temperature,~\cite{wu2019room} a fundamental challenge arises when transferring these concepts to valleytronics. The chiral asymmetry permits only coupling to interacting objects of the same handedness (e.g. chiral molecules). In 2D-TMDs, however, the handedness of chiral valley-excitons is defined by the spin-angular momentum of excitation~\cite{Caruso2022chirality} which can take two possible states ($\pm$). Consequently, nanostructures with equal responses to a valley-excitonic state and its mirror image, namely achiral nanostructures, emerge as favorable choice for valley-based information processing.\\    
    For instance, integrating 2D-TMDs with achiral dielectric metasurfaces demonstrated potential in controlling directionality, lifetime and spectral shape of the PL response.~\cite{bucher2019tailoring,liu2023controlling} Importantly, Liu and coworkers further demonstrated that the DOCP can be enhanced (equally for $\sigma^+$- and $\sigma^-$-polarized excitation) using Mie-resonant metasurfaces.~\cite{liu2023controlling} Other nanophotonic structures facilitated the generation of valley-polarized plasmon/photon–exciton polaritons opening new ways for valley control.~\cite{chervy2018room,vadia2023magneto} Of crucial importance for the development of on-chip valleytronic devices is directional routing of the valley-pseudospin information. Routing has been reported in structures supporting guided modes with spin-momentum locking~\cite{chervy2018room,yang2019chiral,gong2018nanoscale,sun2019separation} and photonic crystals.~\cite{rong2020photonic,wang2020routing} Despite these advancements, an apparent lack of experimental progress remains in the manipulation of valley-selective emission at the level of single nanoantennas. A notable divergence is observed in this context: The experimentally observed effects, though valuable, fell short of the expected efficiency in numerical simulations.~\cite{wen2020steering} Additionally, other nanophotonic designs require the employment of electron beam excitation techniques,~\cite{zheng2021deep,zheng2023electron} which significantly increases the technical complexity as well as the costs of the proposed schemes. Demonstrating efficient schemes for nanoantenna-based valley-routing employing widely accessible and integrated optical techniques therefore remains an open challenge.
    As the investigated nanophotonic architectures gain in complexity, the crucial prerequisite of precisely modelling the electromagnetic interaction between the valley-selective emitters and the resonant modes of these nanostructures becomes challenging and poses a limitation for further developments in this field. In light of this, we investigate a simplified model system as sketched in \autoref{fig:concept}. Namely, we focus on a spherically symmetric (and hence achiral) gold nanoparticle (GNP) resonantly interacting with valley-specific emission from a monolayer of molybdenum disulfide (\ML{}). By performing polarization-resolved cryogenic PL microscopy, we study the farfield polarization properties of this hybrid system under circularly-polarized optical pumping. Although it seems intuitive that an achiral nanoantenna should not affect the valley-pseudospin as it equally interacts with both K and K' valleys, we observe robust quenching of the DOCP in the farfield. Along with that there is no significant increase in the linear polarization components caused, for example, by the ellipticity of the GNP. This leads us to the conclusion that the observed effect represents a predominating depolarization.
    This phenomenon raises fundamental questions about the underlying mechanisms of polarization effects within such hybrid systems. Specifically, what causes the observed depolarization? Is this a result of near-field effects during the excitation phase, or does it occur during the re-emission of photons? Furthermore, which valleys are actually excited in the process? These are the key questions we aim to address in our investigation.\\
    We compare our findings with a systematic numerical analysis of both excitation and emission aspects of the conducted experiments. We thereby isolate the main mechanisms that govern the observed depolarization. In particular, we find that the degree of circular polarization drops dramatically once the excitons are positioned only a few tens of nanometers away from the symmetry point leading to complete depolarization after averaging over contributions from excitons within the optical resolution limit. With this work, we not only aim to refine the existing simulation approaches for valleytronic devices, but also contribute to the deeper understanding of the rich physics of light-valley interactions at the nanoscale.
\section{Results}
\subsection{Sample preparation}
\begin{figure}[h]
        \centering
        \includegraphics[width=8cm]{"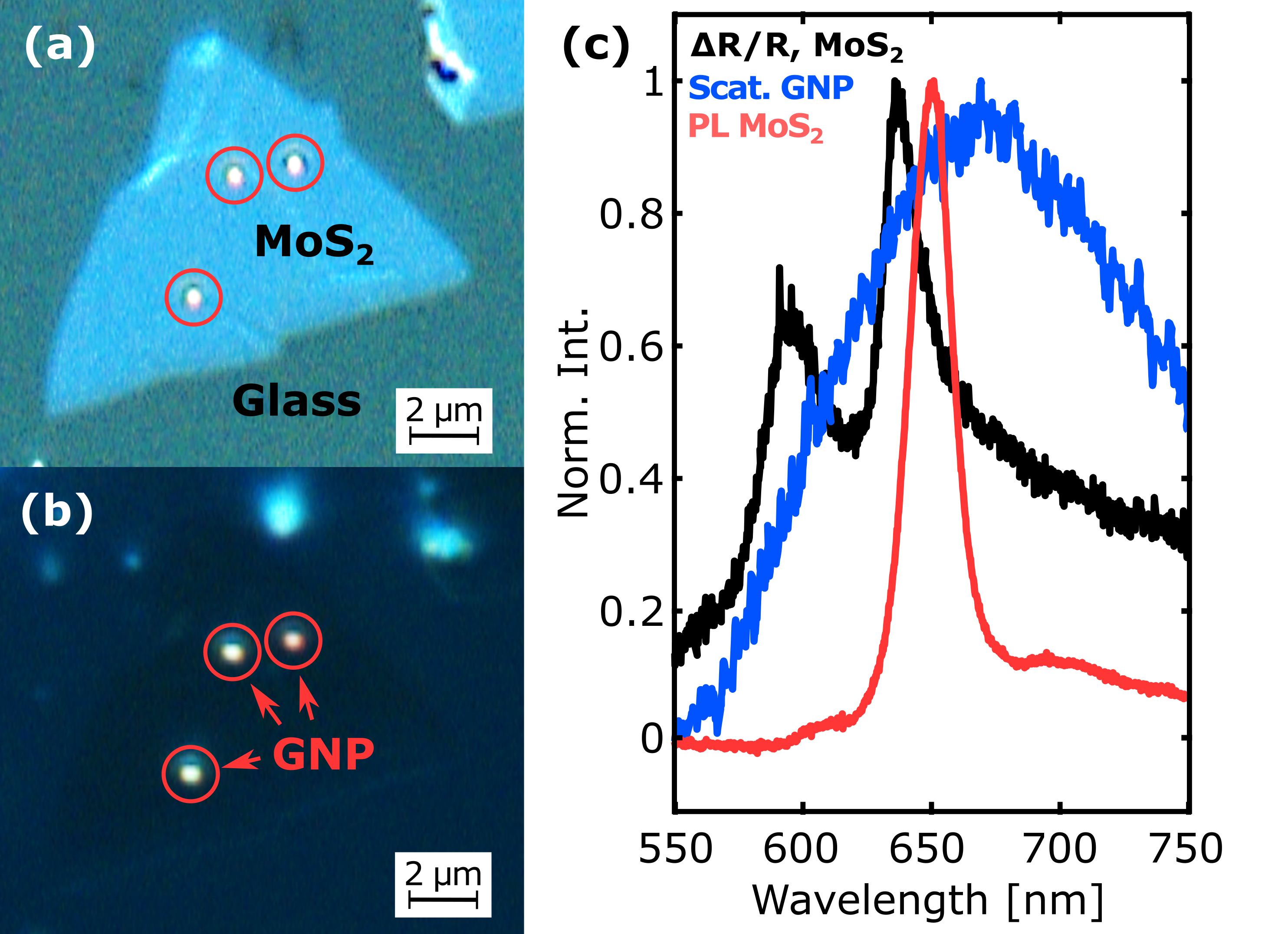"}
		{\phantomsubcaption\label{fig:sample_a}}
		{\phantomsubcaption\label{fig:sample_b}}
		{\phantomsubcaption\label{fig:sample_c}}
		\caption{
            \textbf{Optical microscopy and spectroscopy pre-characterization.} (a) Optical brightfield and (b) darkfield microscope image of a prepared substrate incorporating embedded \ML{} crystals and being decorated by several GNPs. The red circles indicate the positions of GNPs lying atop an embedded \ML{} crystal. (c) Measured cryogenic ($T=\SI{3.8}{\kelvin}$) differential reflectance $\Delta R/R$ (black curve) and PL (red curve) spectrum of an embedded \ML{} crystal as well as the white light reflection spectrum of an isolated GNP (blue curve).}\label{fig:sample}
	\end{figure}

    We prepared hybrid nanoparticle-on-substrate structures incorporating embedded \ML{} by means of a simple spin-coating scheme (see Methods for a detailed description of all processes). Initially, we synthesized \ML{} by chemical vapour deposition (CVD) on silicon/silicon-dioxide wafers~\cite{george2019controlled}. The growth process yields a dense coverage of the substrate with high-optical quality \ML{} crystals~\cite{shree2019_high} reaching edge lengths of up to \SI{60}{\micro\metre}. Next, we transferred the as-grown \ML{} crystals onto a glass wafer by polymethylmethacrylate assisted wet-transfer~\cite{Winter2018_lateral}. Subsequently, we coated the sample with \SI{15}{\nano\metre} silicon oxide using an optimized physical vapour deposition process~\cite{knopf2019integration}. The dielectric spacer layer prevents charge-transfer induced quenching of emission from \ML{} by avoiding direct contact with the GNPs. Finally, by spin-coating we sparsely distributed monodispersed GNPs with an average size of \SI{220 \pm 15}{\nano\metre} on top of the prepared substrates. \autoref{fig:sample_a} and \autoref{fig:sample_b} show an optical bright-field and dark-field microscope image, respectively, of a typical sample. The embedded \ML{} crystals are decorated by several monodispersed GNPs which are visible in both images as bright spots as indicated by circles.
    Next, we characterized the optical properties of the prepared sample at cryogenic temperature as shown in \autoref{fig:sample_c} (see Methods). The differential reflectance $\Delta R/R$ spectrum (black curve) of the embedded \ML{}, which was measured with a tungsten-halogen white light source, shows two distinct peaks at \SI{645}{\nano\metre} and \SI{595}{\nano\metre} wavelength which are related to the A- and B-excitonic resonances formed at the direct bandgap in the K/K' points of the Brillouin zone. The pronounced absorption peaks indicate a large oscillator strength of the A- and B-excitons formed in our samples showing the high-optical quality of the embedded \ML{}. In this study, we focus on the emission from the embedded \ML{} which is dominated by the lower-energetic A-exciton as shown in the PL spectrum as measured from the same sample (red curve). The observed PL peak centered at \SI{650}{\nano\metre} wavelength is slightly Stokes-shifted with respect to the A-exciton absorption peak and has a line width of about \SI{60}{\milli\eV} which is comparable to the values of as-grown \ML{} from our previous work~\cite{shree2019_high} indicating the non-degradative character of the silicon oxide deposition process. We further verified the resonant character of the deposited GNPs at the central emission wavelength of the embedded \ML{} crystals by measuring its white light reflectivity spectrum (blue curve). The GNP exhibits an electric dipolar resonance visible as broad peak centered at \SI{670}{\nano\metre} wavelength. Despite a slight red-shift of the GNP reflectivity spectrum with respect to the A-exciton energy of the embedded \ML{} crystals, the broad width of the plasmonic resonance provides sufficient spectral overlap.
\subsection{Polarization-resolved cryo-PL measurements}
\begin{figure}[h]
		\centering
		\includegraphics[width=16cm]{"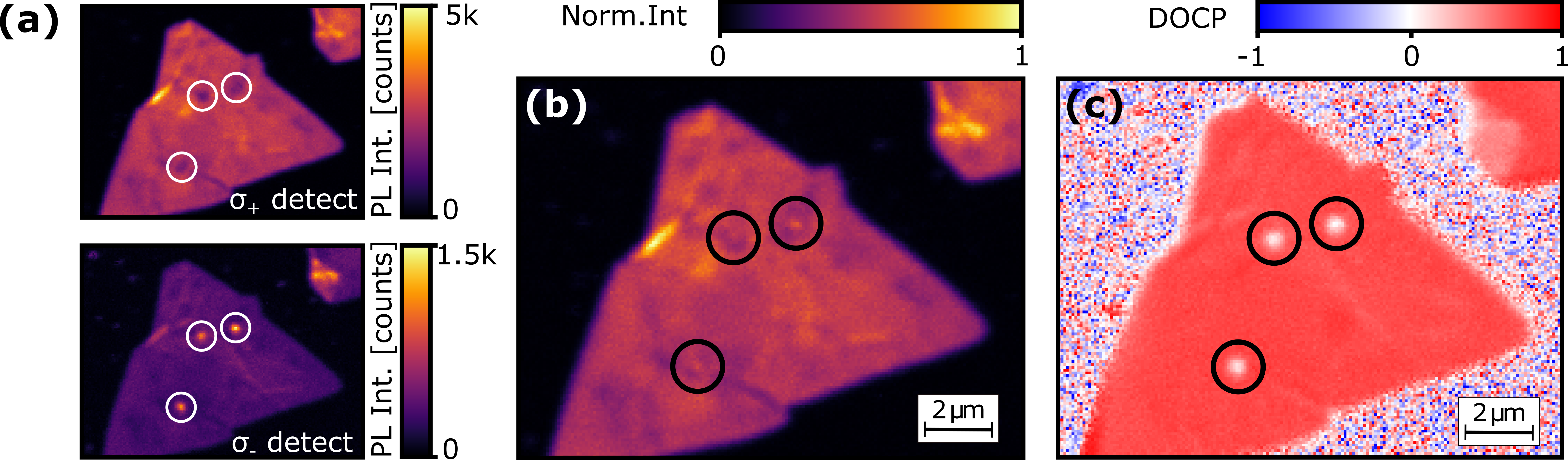"}
		{\phantomsubcaption\label{fig:plresults_a}}
		{\phantomsubcaption\label{fig:plresults_b}}
		{\phantomsubcaption\label{fig:plresults_c}}
		\caption{
            \textbf{Polarization-resolved photoluminescence microscopy.} (a) Measured confocal scans of valley-specific emission intensities $\mathcal{I}_{\sigma^\pm}$ from embedded \ML{} decorated with monodispersed GNPs upon $\sigma^+$ excitation and collected through a \SI{660}{\nano\metre} bandpass filter. (b) Total intensity $\mathcal{I}_\text{tot}=\left(\mathcal{I}_{\sigma^+}+\mathcal{I}_{\sigma^-}\right)$ and (c) degree of circular polarization $\mathrm{DOCP}=\left(\mathcal{I}_{\sigma^+} - \mathcal{I}_{\sigma^-}\right)/\mathcal{I}_\text{tot}$ scans as calculated from the results shown in (a).
		}\label{fig:plresults}
	\end{figure}

    We investigated the modification of the farfield degree of circular polarization (DOCP) of valley-specific emission from \ML{} when scattered by a resonant GNP by performing polarization-resolved PL imaging at cryogenic temperature $T=\SI{3.8 \pm 0.1}{\kelvin}$. To achieve valley-selective excitation of excitons, we pumped the sample near-resonantly at $\SI{633}{\nano\metre}$ wavelength and with $\sigma^+$ polarization. For excitation we used a 100x/0.9 NA objective and an average laser power of 50 µW (16 kW/cm$^2$ peak intensity). In detection, light was collected in reflection geometry using the same objective and a (660 ± 5) nm wavelength bandpass. We further analyzed the collected light in a helical basis ($\sigma^\pm$) to obtain the valley-specific emission intensities ($\mathcal{I}_{\sigma^\pm}$) and calculated the farfield DOCP as $\mathrm{DOCP}=\left(\mathcal{I}_{\sigma^+} - \mathcal{I}_{\sigma^-}\right)/{\mathcal{I}_\text{tot}}$ where ${\mathcal{I}_\text{tot}}=\left(\mathcal{I}_{\sigma^+} + \mathcal{I}_{\sigma^-}\right)$ is the total emission intensity. We have detailed the polarization control and notation of our experimental setup in Sec.~S1 of the Supporting Information. In this notation, the DOCP of the incoming excitation laser and the measured PL emission will have equal sign. In order to investigate the modification of the farfield DOCP of emission from embedded \ML{} by the resonant GNP, we measured $\mathcal{I}_{\sigma^\pm}$ as a function of position by confocal scanning microscopy as shown in \autoref{fig:plresults_a}.
    For both $\mathcal{I}_{\sigma^+}$ and $\mathcal{I}_{\sigma^-}$, we observe a uniform distribution across the embedded \ML{} crystal area where $\mathcal{I}_{\sigma^+}$ is about 6-times higher than $\mathcal{I}_{\sigma^-}$ as a signature of the induced exciton valley-polarization due to the valley-selective excitation. A different situation, however, is observed at the positions of the GNPs which are highlighted by circles. We find a significant local modulation of the valley-specific emission intensities with $\mathcal{I}_{\sigma^+}$ being slightly reduced and $\mathcal{I}_{\sigma^-}$ being noticeably enhanced with respect to the case without GNPs. To further quantify this effect we calculated $\mathcal{I}_{\text{tot}}$ and DOCP as shown in \autoref{fig:plresults_b} and \autoref{fig:plresults_c}, respectively. While there is only a small modulation of $\mathcal{I}_{\text{tot}}$ by the GNPs, we observe a strong reduction $\Delta_{\text{DOCP}}=0.63\pm 0.11$ from $0.71\pm 0.03$ for embedded \ML{} without GNP to $0.08\pm 0.08$ with GNP. As confirmed by full-polarization resolved measurements, the emission from the embedded \ML{} with GNP exhibits negligible linear polarization components (see Sec.~S2 of the Supporting Information). Hence, the observed reduction in the DOCP is equivalent to a reduction of the total degree of polarization which can be clearly attributed to the presence of the GNPs. Interestingly, the cylindrical symmetry of the nanoparticle-on-substrate geometry considered in this work would suggest a preserved DOCP of emission which is in stark contrast to our experimental observations. In the following, we systematically analyze aspects of excitation and emission in our hybrid system and compare them with predictions from numerical simulations which allows us to isolate the mechanism leading to the observed reduction in the DOCP.
\subsection{Nearfield excitation polarization}
    The initial requirement to observe valley-specific emission from \ML{} is a valley-selective excitation. By selection rules, the excitation rate of carriers in valleys K and K' is proportional to the intensity of the $\sigma^{+}$ and $\sigma^{-}$ polarized components of the external field, respectively. In \ML{}, the out-of-plane contributions from spin-forbidden dark excitons are negligible without strong external magnetic fields as shown by Robert et al.~\cite{Robert2020measurement} Therefore, in equilibrium the local exciton densities $n_K(x,y)$ and $n_{K'}(x,y)$ will be proportional to $I^{||}_{\sigma^{+}}(x,y)$ and $I^{||}_{\sigma^{-}}(x,y)$, respectively, where the superscript $||$ refers to the in-plane components of the external field. Note that, throughout this manuscript, $I$ quantifies nearfield intensities, while $\mathcal{I}$ relates to farfield intensities. The induced degree of valley-polarization, $\eta = \left(n_K-n_{K'}\right)/\left(n_K+n_{K'}\right)$, is therefore proportional to the 2D-DOCP, i.e. the DOCP of the in-plane components of the excitation field. In order to analyze the influence of the GNP on the valley-selective excitation of \ML{}, we numerically calculate the helical intensities, $I^{||}_{\sigma^{\pm}}\propto|E^{||}_{\sigma^{\pm}}|^2$, in a plane \SI{15}{\nano\metre} below the GNP for a $\sigma^+$-polarized excitation beam.
    \begin{figure}[h]
        \centering
        \includegraphics[width=16cm]{"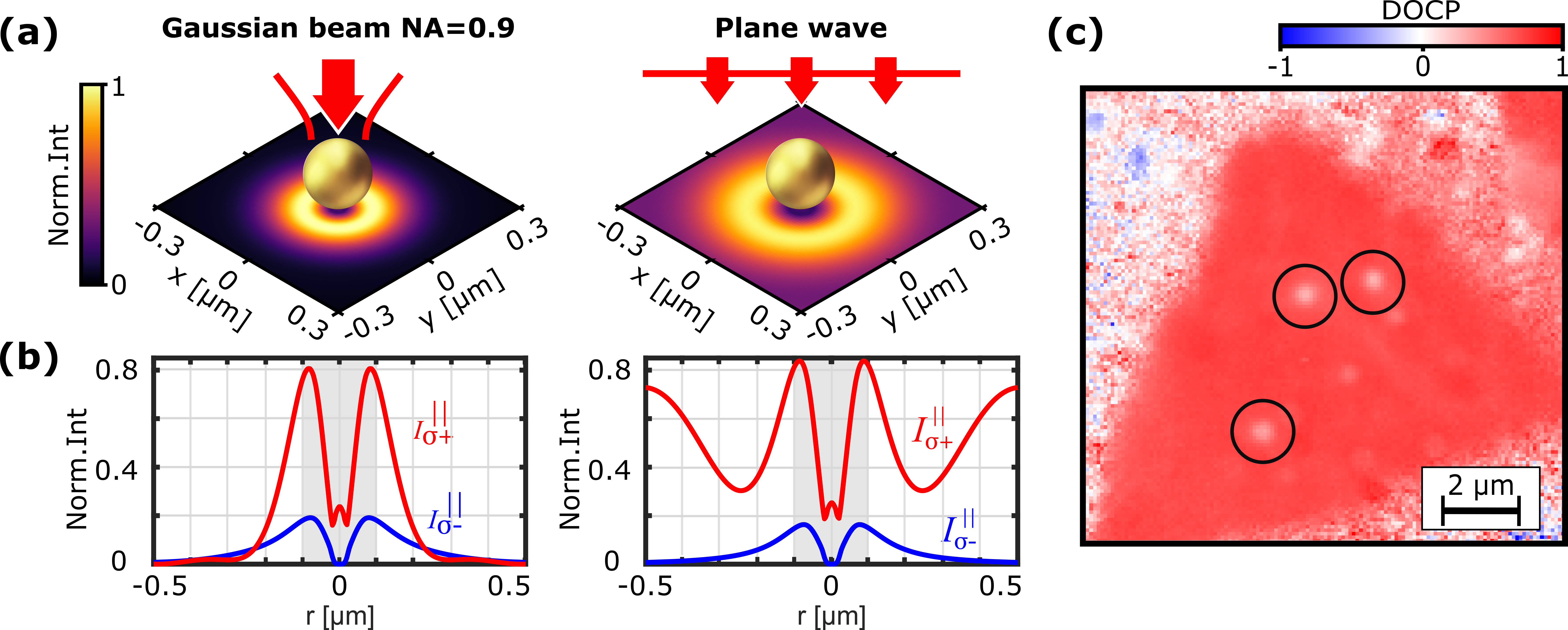"}
        {\phantomsubcaption\label{fig:excitation_a}}
        {\phantomsubcaption\label{fig:excitation_b}
        \phantomsubcaption\label{fig:excitation_c}}
        \caption{ 
            \textbf{Impact of the gold nanoparticle on the valley-selective excitation of \ML{}.} (a) Total intensity of the external in-plane nearfield in the plane of the monolayer upon $\sigma^{+}$ polarized excitation with a focused Gaussian beam (NA=0.9) positioned at the center of the GNP (left) and a plane wave (right). (b) Respective individual contributions of $\sigma^+$ and $\sigma^-$ polarized components in dependence on the distance $r=\sqrt{x^2+y^2}$. The normalization is chosen such that the sum of both curves peaks to 1. Note that identical results are obtained for $\sigma^{-}$ polarized excitation up to exchanged labels $\sigma^{\pm}$, respectively. (c) Measured DOCP of valley-specific emission from embedded \ML{} decorated with monodispersed GNPs upon $\sigma^+$-polarized wide-field illumination and collected through a \SI{660}{\nano\metre} bandpass filter.
        }\label{fig:excitation}
    \end{figure}
    \autoref{fig:excitation_a} shows the resulting total in-plane intensities generated by a focused Gaussian beam (left) and a plane wave (right). In both cases, the maximum intensity is concentrated in a ring around the GNP. We note that this pattern is characteristic when observing closely to a spherical GNP, within distances less than 50 nm, where near-field interactions with nanoantennas are typically prominent. Further insights into the formation of such a ring can be found in Sec.~S5 of the Supporting Information. The individual contributions of $\sigma^+$ and $\sigma^-$ polarized field components to these intensities are plotted in \autoref{fig:excitation_b}, represented by red and blue curves respectively, against the distance $r$ from the GNP's center. For both excitation conditions, the GNP induces the cross-polarized intensity component $I^{||}_{\sigma^{-}}$ peaking at the projected GNP's edge (gray shaded region) and diminishing with increasing distance. Notably, this cross-polarized component is slightly larger for the focused beam. Reference simulations of a tightly focused Gaussian beam without a GNP (see Fig.~S6 of the Supporting Information) reveal that $I^{||}_{\sigma^{-}}$ remains nearly zero, emphasizing that the slight increase of this component observed in the presence of a GNP is not due to high NA itself. Additionally, for the focused beam the intensity of $\sigma^+$-polarized field drops faster than $\sigma^-$, resulting in their curves intersecting at $|r|\approx$\SI{0.22}{\micro\metre}. In contrast for a plane wave, the total field at larger distances from the GNP is dominated by the incident field characterized by pure $\sigma^+$ polarization. The resultant 2D-DOCP maps for all of the discussed cases are provided in Sec. S6 of the Supplementary Information.


    Hence, the excitation scheme is expected to influence the observed reduction in DOCP of emission from \ML{} mediated by the GNP. We therefore repeated our cryo-PL imaging experiments using wide-field illumination (excitation-NA$\,\approx 0$) mimicking the plane wave excitation. \autoref{fig:excitation_b} shows the respective measured DOCP image of PL from the same \ML{} sample as shown before and measured with an average excitation power of 200 µW (40 W/cm$^2$ peak intensity). Indeed, we observe a less pronounced local reduction of $\Delta_{\text{DOCP}}=0.39\pm 0.06$ from $0.72\pm 0.02$ for embedded \ML{} without GNP to $0.33\pm 0.04$ with GNP. While the lower $\Delta_{\text{DOCP}}$ is qualitatively in line with the prediction from our nearfield simulations, the observed difference is, however, still significant. For a quantitative comparison of both excitation schemes on the basis of our numerical simulations, we need to take into account the finite optical resolution in our experiments. In case of the wide field scheme, we use a Gaussian point spread function to obtain the averaged helical excitation intensities $\tilde{I}_{\sigma^\pm}(x,y)\propto\iint G(\xi-x,\eta-y)I^{||}_{\sigma^\pm}(\xi,\eta)\text{d}\xi\text{d}\eta$, where $G(x,y)$ is a Gaussian weight function given by \autoref{eq:Gaussian} from the Methods section. For the confocal scanning scheme the resolution limit is encoded in the finite beam size. Importantly, during scanning the beam position relative to the GNP center changes. Therefore, the averaged helical excitation intensities at a distance $r$ from the center of the GNP can be estimated as $\tilde{I}_{\sigma^\pm}(r)\propto\iint I_{\sigma^{\pm}}^{||}(r,\xi,\eta)\text{d}\xi\text{d}\eta$, where $I_{\sigma^{\pm}}^{||}(r,x,y)$ are the respective nearfield intensities resulting from the excitation with the focused beam that is displaced by the distance  $r$ from the GNP center. Finally, the numerically predicted farfield PL DOCP for each measurement setting is computed using \autoref{eq:DOCP}. We obtain a predicted reduction of $\Delta_{\text{DOCP}}=0.18\pm 0.04$, i.e. from $0.72\pm 0.02$ (measured) without to $0.54\pm 0.02$ (calculated) with GNP, and $\Delta_{\text{DOCP}}=0.45\pm 0.06$, i.e. from $0.71\pm 0.03$ (measured) without to $0.26\pm 0.03$ (calculated) with GNP, for the plane wave and Gaussian beam excitation, respectively. Our calculations based on excitation effects are consistent with the experimental observation of a larger reduction in DOCP for the case of the confocal as compared to the widefield excitation scheme. However, the experimentally observed changes of $\Delta_{\text{DOCP}}=0.39\pm 0.06$ and $\Delta_{\text{DOCP}}=0.63\pm 0.11$ remain systematically larger indicating additional contributing factors.\\
    Note, that in this simple estimation we have not assumed any exciton diffusion which is known to potentially impact the electromagnetic interaction of emitters in 2D-TMDs with optical nanoresonators~\cite{Raziman2022_exciton}. In our system, however, we expect a rather weak impact of the nearfield of the GNP on the diffusion properties of the excitons in \ML{} as first, we observe no significant modulation of the total cryo-PL intensity at the positions of the GNPs, and second, we find the depolarization effect to be robust across all our samples including different spacer thicknesses (see Sec.~S4 of the Supporting Information) and GNPs with naturally varying size and topography. Nevertheless, the observed DOCP of emission from \ML{} with GNP cannot be explained by excitation effects alone but additional factors need to be considered as discussed next.
\subsection{Farfield emission polarization}
\begin{figure}[h]
        \centering
        \includegraphics[width=16cm]{"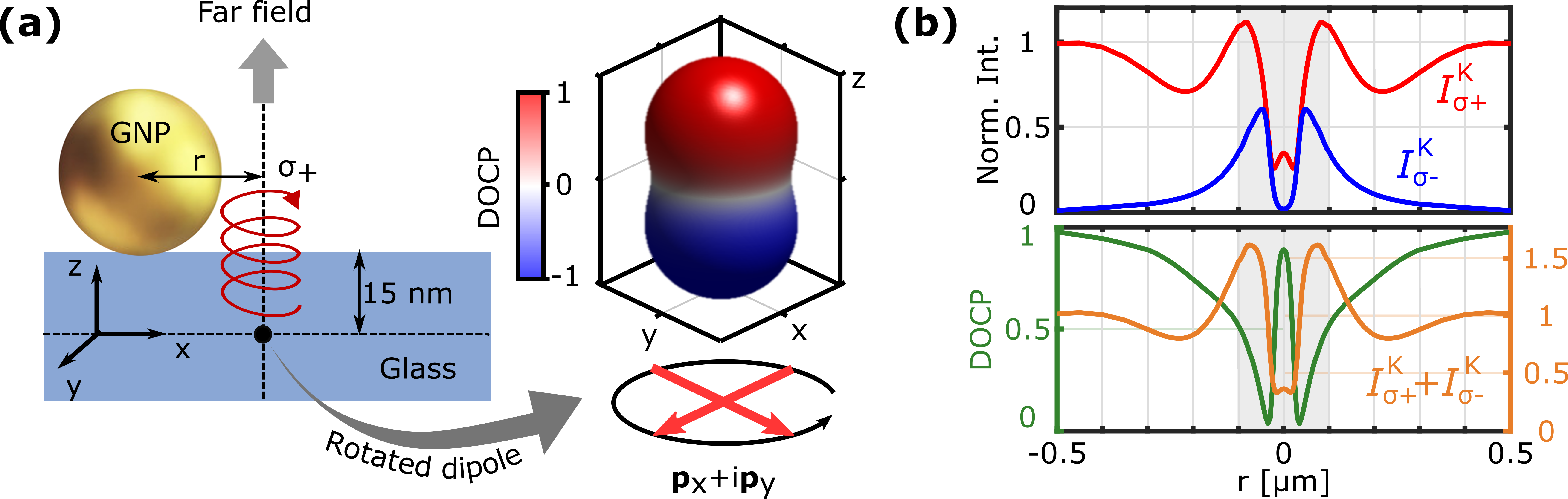"}
	{\phantomsubcaption\label{fig:emission_a}}
	{\phantomsubcaption\label{fig:emission_b}}
	\caption{
            \textbf{Impact of the gold nanoparticle on the photoluminescence polarization from distributed emitters.} (a) Sketch of the simulated nanoparticle-on-substrate geometry showing the position of the rotating electric dipole in a plane \SI{15}{\nano\metre} below and displaced by a distance $r$ from the projected center of the GNP. The inset shows the calculated farfield radiation pattern of a rotating electric dipole $\Vec{\mathbf{p}}_{K}$, the color encodes the respective DOCP. (b) Top: Calculated farfield intensities $\mathcal{I}^{K}_{\sigma^{+}}$ (red line) and $\mathcal{I}^{K}_{\sigma^{-}}$ (blue line) emitted by a rotating dipole $\Vec{\mathbf{p}}_{K}$ as a function of the displacement distance $r$. Bottom: The corresponding DOCP (green curve) and total intensity (orange curve) of the integrated farfield. The farfield intensity is normalized to the intensity obtained without a GNP. Note that the results for a dipole with opposite spin, $\Vec{\mathbf{p}}_{K'}$, can be obtained by exchanging the labels $\sigma^{\pm}$, respectively. 
        }\label{fig:emission}
	\end{figure}

    By symmetry, the nanoparticle-on-substrate geometry is expected to preserve the circular polarization of valley-selective emission in the forward scattered farfield. This can be understood by modelling emission using dipoles exhibiting circularly polarized farfield components. For this purpose, different types of emitters are discussed in literature with focus on their symmetry properties~\cite{Cameron2012_optical, fernandez2013role, Lamprianidis2022_directional} or different multipolar coupling behaviour.~\cite{zambrana2016tailoring, eismann2018exciting} The prevalent model used in the context of 2D-TMDs is the rotating electric dipole,~\cite{gong2018nanoscale, chervy2018room, li2018tailoring, sun2019separation, hu2019coherent, liu2023controlling, zheng2023electron} $\Vec{\mathbf{p}}_{\sfrac{K}{K'}}=\Vec{\mathbf{p}}_x~\sfrac{+}{-} ~i\Vec{\mathbf{p}}_y$, where the spin or valley index is associated with the fixed phase sign. On the right side of \autoref{fig:emission_a} we show the farfield intensity distribution and respective DOCP of a single counterclockwise rotating electric dipole. Due to its fixed axis of rotation, $\Vec{\mathbf{p}}_{K}$ emits circularly polarized light with opposite handedness into different half spaces matching the PL polarization properties of valley-selective excitons in bare 1L-TMDs (see Sec.~S3 of the Supporting Information). 
    \noindent When placed below a resonant GNP, the rotating electric dipole is expected to induce a respective mirror rotating dipole with equal spin (as both linear dipole components of the mirrored rotating dipole will experience the same $\pi$ phase flip). Consequently, the PL polarization properties of the hybrid system would be similar to that of a bare rotating electric dipole. However, this is only the case for a single rotating dipole placed on the symmetry axis of the hybrid system. For a finite displacement $r>0$ of the rotating dipole from the projected center of GNP, the cylindrical symmetry of the system is broken. To investigate the effect of the dipole displacement $r$ on the farfield polarization of the hybrid system, we calculate the resulting farfield distribution for different positions of $\Vec{\mathbf{p}}_{K}$ (see Methods). By integration over a numerical aperture of 0.9, we obtain the helical farfield intensities $\mathcal{I}^K_{\sigma^\pm}(r)$ as plotted on top of \autoref{fig:emission_b}. Similar to the previously analyzed nearfields, we observe the emergence of the cross-polarized intensity component $\mathcal{I}^K_{\sigma^-}(r)$, peaking at about $|r|\approx60$ nm, with its contribution being notably pronounced. Below, in \autoref{fig:emission_b} we show the resulting total farfield intensity (orange curve) and the farfield DOCP (green curve) as functions of the displacement distance $r$. Here, all the farfield intensities are normalized to the emission of a rotating electric dipole without GNP, i.e. at infinite distance. 
    As expected from symmetry, we obtain a high farfield DOCP for dipoles below the center of the GNP ($r=0$). However, we find that the DOCP reduces radically even for small displacements below the footprint of the GNP (shaded region) and reaches a minimum of about 0.04 for $r=\SI{35}{\nano\metre}$. Simultaneously, the relative intensity contribution of dipoles below the center of the GNP is significantly lower than for emitters located below the edge of the GNP. Our findings therefore highlight that the overall farfield polarization of the hybrid system with an ensemble of spatially separated valley excitons can differ significantly from the intuition based on a single rotating electric dipole below the center of the GNP.
\subsection{Full model for optically addressing the valley degree of freedom}
\begin{figure}[h]
        \centering
        \includegraphics[width=8cm]{"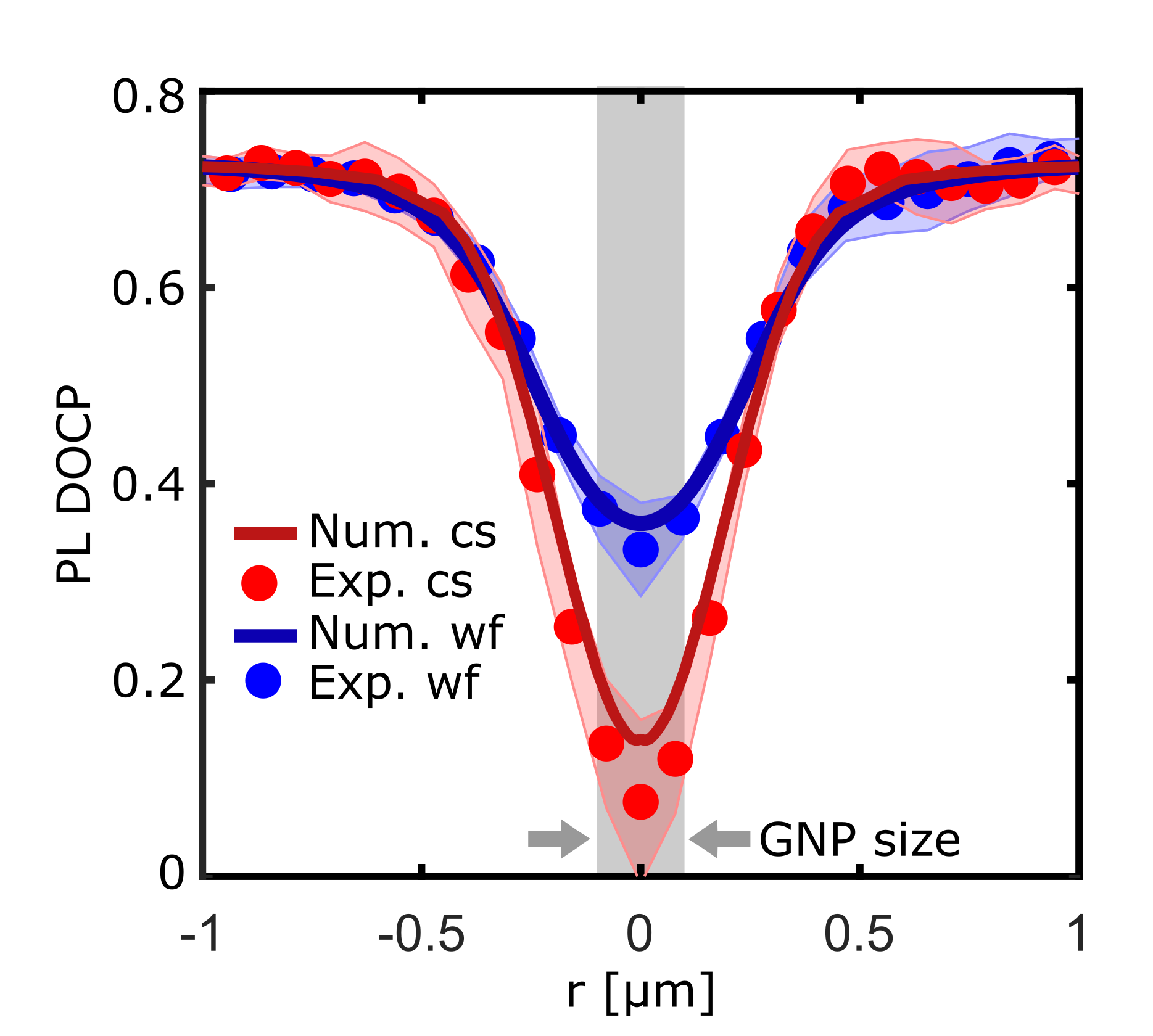"}
        \caption{
            \textbf{Quantitative modelling of the collective photoluminescence polarization.} Comparison of the spatial lines profiles of the calculated (solid curves) and measured (circles) DOCP as a function of the radial distance $r$. Blue and red color corresponds to wide field (wf) and confocal scanning (cs) measurement settings, respectively.
        }\label{fig:excitation&emission}
    \end{figure}

    In this section, we aim to provide a quantitative model that describes the observable PL intensity and polarization of the hybrid system in the farfield by combining excitation and emission effects as well as incorporating the finite optical resolution. As excitonic emitters in 2D-TMDs are created across the whole excitation area, their distribution needs to be taken into account. We model the emission from such an ensemble of excitons by averaging over a distribution of rotating electric dipoles $\Vec{\mathbf{p}}_{K/K'}(x,y)$ for which the local exciton densities $n_{K/K'}(x,y)$ are proportional to the helical excitation intensities $I^{||}_{\sigma^{\pm}}(x,y)$ below the GNP. Note that for the hybrid system an excitation beam with well-defined $\sigma^+$ polarization leads to non-zero exciton densities for both valleys $n_{K}(x,y)$ and $n_{K'}(x,y)$. Simultaneously, an exciton located in the valley K or K' leads to non-zero farfield intensities of both helicites which we denote as $\mathcal{I}^{K}_{\sigma^{\pm}}$ or $\mathcal{I}^{K'}_{\sigma^{\pm}}$, respectively. Then, we describe the helical farfield intensities observed for the hybrid system in wide-field illumination according to Equation 2 and using Gaussian smoothing (where the Gaussian width matches the optical resolution of the experiments, see Methods section for further details). Similarly, we calculate the result for the case of confocal scanning according to Equation 4 where we take into account the dependence of the excitation field on the center position of the focused excitation beam as previously discussed. For the hybrid system, we recall that a $\sigma^+$ polarized excitation beam leads to non-zero exciton densities in both valleys K/K' proportional to the in-plane intensity distribution of the total field, $n_{\sfrac{K}{K'}}\propto I^{||}_{\sigma^{\sfrac{+}{-}}}$. Similarly, we have discussed above that a single exciton located in the valley K (or K') leads to both $\sigma^{+}$ and $\sigma^{-}$ polarized PL intensities in the farfield which we denoted as $\mathcal{I}^{K}_{\sigma^{\pm}}$ (or $\mathcal{I}^{K'}_{\sigma^{\pm}}$), respectively, where the farfield intensities from emitters in opposite valleys are obtained as $\mathcal{I}^{K'}_{\sigma^{\pm}} = \mathcal{I}^{K}_{\sigma^{\mp}}$. Assuming a fixed excitation beam (as valid for widefield excitation), we can express the farfield PL contributions emitted from a point (x,y) of the \ML{} crystal in a helical basis as $\mathcal{I}_{\sigma^\pm}(x,y)=n_{K}(x,y)\cdot\mathcal{I}^{K}_{\sigma^\pm}(x,y)+n_{K'}(x,y)\cdot\mathcal{I}^{K'}_{\sigma^\pm}(x,y)$. In order to obtain the farfield observable distribution we convolute the local contributions $\mathcal{I}_{\sigma^\pm}(x,y)$ with a Gaussian point spread function where the Gaussian width matches the optical resolution of the experimental setup (see~\autoref{eq:Gaussian} in Methods). For the case of confocal microscopy, the excitation beam is not fixed but instead scanned across the sample. Hence, the exciton density distribution has to be calculated for each excitation beam position separately and the farfield observable PL contributions are then described by~\autoref{eq:GassianSmoothing} as shown in the Methods section.\\
    Ultimately, we compare the calculated farfield DOCP with our measured results for both excitation schemes in \autoref{fig:excitation&emission}.
    Here, we plotted the cross-sections of the measured DOCP of emission from \ML{} upon $\sigma^+$ excitation across the GNPs both for wide-field illumination (blue circles) and confocal scanning (red circles). The cross-sections and error corridors (shaded regions) are obtained by averaging over several GNPs as well as cross-section directions across the GNP.
    The cross-sections and error corridors (shaded regions) are computed as the mean value and the standard deviation, respectively. For our sampling we have used the DOCP shown in \autoref{fig:plresults_c} (scanning) or \autoref{fig:excitation_b} (wide field) and compared the cross-sections along x- and y- axes of three measured nanoparticles on a single flake (in total 6 samples per measurement scheme)
    For comparison, we also plotted the numerical prediction of the farfield DOCP according to \autoref{eq:DOCP} (see Methods) for plane wave excitation (blue solid curve) as well as for confocal Gaussian beam excitation (red solid curve). Our numerical calculations closely agree with the experimental results indicating that the presented model accurately incorporates the dominating physical effects leading to the observed reduction of the DOCP of emission from \ML{}.\\
    Interestingly, similar depolarization effects were also described in a very recent theoretical study by Salzwedel et al.~\cite{salzwedel2023_spatial} for strongly coupled plexcitonic systems on the basis of dipole-dipole interaction. The described systems in their work exhibiting different coupling regimes can serve as further test bed for the validity of our proposed model.
\section{Conclusion}
    In this work, we investigated the complex polarization behaviour of nanoscopic hybrid systems consisting of 2D-TMDs interacting with resonant nanostructures. For our hybrid model system, we used a spherical GNP as a probe to systematically investigate its impact on the emission polarization behaviour of a 2D-TMD crystal below. Specifically, we fabricated CVD-grown \ML{} embedded within amorphous silicon oxide thin-films being decorated with resonant GNPs and investigated the modification of the DOCP of valley-selective emission from the embedded \ML{} by the presence of monodispersed GNPs. We reported a robust and strong reduction in the DOCP as mediated by the GNPs which is in contrast to the expectation that a cylindrically symmetric system conserves the circular polarization state in the direction out of the substrate.\\
    Our observation highlights the need for a predictive model able to quantitatively describe the collective response of such nanoscopic hybrid systems. Therefore, we further investigated the impact of the GNPs on the measured DOCP by numerically and experimentally analyzing effects at the level of excitation, emission and detection resolution. While we have found the excitation polarization to be strongly affected at high numerical aperture, the impact can be reduced by employing a wide-field excitation scheme mimicking plane wave excitation. Our respective measurements, however, showed that a significant reduction in the DOCP is still observed beyond the prediction according to the excitation effect alone. Next, we numerically analyzed the impact of the spatial extent of the emitting \ML{} leading to emission contributions from out of the symmetry point of the GNP-on-substrate geometry. Our simulations showed a crucial sensitivity of the farfield DOCP on the emitter position even for emitters located well within the footprint area of the GNP. Ultimately, by modelling excitons in valleys K/K' as spatially distributed rotating electric dipoles weighted by their respective excitation rate and averaging contributions from across the optical detection area, we are able to accurately reproduce the experimentally observed reduction in DOCP.\\
    This physical model yields a quantitative description for the farfield PL properties of (ensembles of) valley-selective emitters in 2D-TMDs when in proximity to resonant nanostructures. It gives deeper insights into the complex polarization behaviour of nanoscopic hybrid systems by discerning the individual contributions discussed above which serves as a crucial prerequisite to accurately design and optimize nanoscopic hybrid systems for valleytronic applications. While our work specifically shows the challenges and limitations for the realization of valleytronic devices on the basis of resonant plasmonic nanostructures, it hints at pathways to circumvent these limitations. These might range from optimizing the valley selective excitation of 2D-TMDs in the vicinity of resonant nanostructures by structured light excitation techniques or exploring alternative material platforms and resonance types for tailoring the collective emission response to nanostructuring of 2D-TMDs for precise emitter position control for the realization of functional nanoscopic valleytronic devices.
\section{Methods}
\subsection{CVD growth and embedding of \ML{}}
    Single crystals of \ML{} with high optical quality were synthesized on silicon wafers with 300\,nm thermal oxide (Sil'tronix,  root mean square (RMS) roughness <0.2\,nm) by a modified chemical vapor deposition process in which Knudsen cells were employed for the delivery of precursors.~\cite{george2019controlled, shree2019_high} The grown \ML{} single crystals were characterized initially using optical microscopy and Raman spectroscopy and transferred onto a glass substrate using a poly(methyl methacrylate) assisted wet-transfer process.~\cite{Winter2018_lateral} Subsequently, the sample was coated by a thin film of silicon oxide using an adapted physical vapour deposition process~\cite{knopf2019integration} (Buhler SyrusPro 1100) using a relatively low deposition rate of 0.5\,nm/s and 15\,nm target thickness.
\subsection{Gold nanoparticle deposition}
    Initially, the purchased GNP suspension (Merck/Sigma Aldrich, particle size 200\,nm) was  ultrasonicated for 5\,mins and then 20\,$\upmu$L of the suspension were diluted with 60\,$\upmu$L of isopropanol to reach a particle concentration of $\sim4.7\times10^8$ particles/mL. Subsequently, the diluted suspension was spin-coated onto the sample at 1000\,rpm with 500\,round/s$^2$ acceleration for 20\,s. This process resulted in a homogeneous distribution of mostly isolated GNPs on the substrate.

\subsection{Optical experiments at cryogenic temperatures}  
    Optical experiments were conducted at cryogenic temperatures ($T=\SI{3.8 \pm 0.1}{\kelvin}$) using a closed-cycle helium cryostation (Montana Instruments s50) with high numerical aperture optical access (100x/0.9NA) in reflection geometry. The non-polarizing 30:70 plate beam splitter was utilized for all our cryogenic measurements. The sketch of our setup as well as the detailed characterization of polarization effects of the utilized optical components are provided in Sec.~S1 of the Supporting Information. For white light spectroscopic measurements, a stabilized tungsten-halogen white light source was used. The differential reflectance spectrum in \autoref{fig:sample_c} was obtained from the reflectance of the coated by silicon oxide \ML{} and the reflectance of the coated bare substrate via 
    \begin{equation}
        \Delta R/R=\frac{R_{\mathrm{MoS_2}}-R_{\mathrm{Subs}}}{R_{\mathrm{Subs}}}.
    \end{equation}
    For photoluminescence spectroscopic measurements, the sample was excited using a \SI{561}{\nano\metre} wavelength continuous-wave diode pumped solid-state laser. For polarization-resolved photoluminescence imaging, the sample was pumped near-resonantly at $\SI{633}{\nano\metre}$ wavelength using a continuous wave helium-neon gas laser. The polarization of the excitation beam was prepared by a linear polarizer and quarter-wave phase plate by monitoring the degree of circular polarization of the collimated excitation beam before entering the objective. In detection, the polarization was analyzed by a combination of super-achromatic quarter-wave phase plate and a linear polarizer. For confocal scanning measurements, the excitation beam with an average power of $\SI{50}{\micro\watt}$ was focused to a diffraction limited spot with a diameter of $2r = 2\lambda/(\mathrm{NA}\cdot\pi)\approx\SI{0.45}{\micro\metre}$ and a peak intensity of $\SI{16}{\kilo\watt\per\centi\metre\squared}$. Lateral scanning was performed using two piezoelectrical nanopositioners moving the sample. This allows for stable conditions of the focused excitation with a fixed degree of circular polarization. The collected emission was imaged onto an EMCCD camera (Andor iXon 897 Ultra) and integrated over an area of 3x3 pixels (=\,48x48\,\SI{}{\micro\metre\squared} on camera chip) which relates to an area of 270x\SI{270}{\nano\metre\squared} in the conjugated sample plane. For wide-field measurements, an additional lens was introduced to focus the excitation beam with an average power of $\SI{200}{\micro\watt}$ onto the back-focal plane of the objective. This results in a beam size of $\approx \SI{25}{\micro\metre}$  on the sample and a peak intensity of $\SI{40}{\watt\per\centi\metre\squared}$. In all imaging experiments, the collected emission was filtered by a 650 nm longpass filter to block the laser light collected from the sample in reflection. An additional 660 nm bandpass filter was used to limit the detection to a spectral band as close to the A-excitonic photoluminescence peak as possible.
\subsection{Numerical simulations}
    For the numerical analysis of our hybrid system, we have used the commercial finite-difference time-domain solver (Lumerical FDTD solutions). Perfectly matched layers were applied at each boundary of the simulation area. The nanoparticle was modelled as a gold sphere with a radius of 100 nm lying on a glass that fills the lower halfspace of the FDTD domain. We did not differentiate between the glass substrate and the capping oxide layer, considering both as a homogeneous medium with the refractive index of n=1.5. The remaining space is filled with air. The presence of the monolayer was neglected in our simulations. The optical properties of the materials were taken from the default material database in Lumerical, namely, Handbook of Optical Constants of Solids I - III by E. Palik.\\
    Focusing on the excitation aspect, we first computed the in-plane nearfield components $E_x(x,y)$ and $E_y(x,y)$ at the position of the monolayer, 15 nm below the nanoparticle. The GNP was illuminated under normal incidence by either a plane wave or a Gaussian beam in a thin lens approximation with NA=0.9. For both excitation sources we have set circular polarization and center wavelength of 633 nm. The mesh override sections with the mesh size of 5 nm were applied for the areas around the beam focus and nanoparticle to achieve finer resolution. In case of a plane wave, we computed the local valley exciton density as $n_{K}(x,y)\propto I^{||}_{\sigma^{+}}(x,y)$ and $n_{K'}(x,y)\propto I^{||}_{\sigma^{-}}(x,y)$, where $I^{||}_{\sigma^{\pm}}\propto|E^{||}_{\sigma^{\pm}}|^2=|E_x\pm iE_y|^2$. To reproduce the result of confocal scanning we have displaced the focal spot of the Gaussian beam from the nanoparticle center and computed the corresponding $n_{K/K'}(r,x,y)$ as a function of the displacement distance $r$.\\
    In the second set of simulations we studied the radiation farfield polarization from a single emitter in \ML{} scattered by the GNP. For that we mimicked an emitter from K or K' valley as a pair of perpendicular electric dipoles with 90° or -90° phase shift, respectively. The simulations were performed for different GNP-emitter distances  $r$ as shown in \autoref{fig:emission_a}.  High sensitivity of the interaction between the GNP and the dipole emission on the relative distance yields high sensitivity to the mesh size. Performing the convergence test, we chose 3 nm mesh size around the dipole and the nanoparticle as a good trade-off between the simulation accuracy and computation time. The nearfield was extracted from a horizontal plane right above the nanoparticle and propagated by 1 m to the farfield using the built-in Lumerical functions.  The angular filter was applied to the resulting field restricting the accepted emission angles to fit the finite numerical aperture of our objective (NA=0.9). Next, we transformed the filtered field given in spherical coordinates $\{E_\theta,E_\phi,E_r\}$ into the helical basis $E_{\sigma^\pm}= E_\theta \pm iE_\phi$. The corresponding intensities are given by $\mathcal{I}^{K}_{\sigma^{\pm}}(r)=|E^{K}_{\sigma^\pm}(r)|^2$, where we have added the valley index $K$ and dependence on the emitter displacement. Note that from the symmetry considerations, it is sufficient to consider just one valley, because the intensity contributions of the opposite one will be  swapped, i.e. $\mathcal{I}^{K'}_{\sigma^-}(r)=\mathcal{I}^{K}_{\sigma^+}(r)$ and $\mathcal{I}^{K'}_{\sigma^+}(r)=\mathcal{I}^{K}_{\sigma^-}(r)$.\\
    Having quantified both, the excitation and emission processes, we combined them to compare with the experimental results. In case of the widefield experiment, the farfield intensities of $\sigma^\pm$-polarized light emitted from every point $(x,y)$ is given by
    \begin{equation}\label{eq:widefield}
        \mathcal{I}_{\sigma^\pm}(x,y)=n_{K}(x,y)\cdot\mathcal{I}^{K}_{\sigma^\pm}(x,y)+n_{K'}(x,y)\cdot\mathcal{I}^{K'}_{\sigma^\pm}(x,y)
    \end{equation}
    Next, we took into account the finite resolution of our optical system by applying the Gaussian point spread function $\tilde{\mathcal{I}}_{\sigma^{\pm}}(x,y)=\iint G(\xi-x,\eta-y) \mathcal{I}_{\sigma^\pm}(\xi,\eta)\text{d}\xi\text{d}\eta$, where
    \begin{equation}\label{eq:Gaussian}
        G(x,y)=\frac{1}{2\pi\Sigma^2}\exp{\left(-\frac{x^2+y^2}{2\Sigma^2}\right)}   
    \end{equation}
    with  $\Sigma=174$~nm, as extracted from the widefield experimental data.\\
    In case of the confocal scanning configuration, the local exciton densities become depended on the position of the excitation beam relative to the GNP. Therefore, the detected $\sigma^\pm$ PL intensities at the distance $r$ from the GNP center becomes
    \begin{equation}~\label{eq:GassianSmoothing}
        \tilde{\mathcal{I}}_{\sigma^\pm}(r)\propto \iint n_{K}(r,\xi,\eta)\cdot\mathcal{I}^{K}_{\sigma^\pm}(\xi,\eta)+n_{K'}(r,\xi,\eta)\cdot\mathcal{I}^{K'}_{\sigma^\pm}(\xi,\eta)\text{d}\xi\text{d}\eta.  
    \end{equation}
    Note, that here there is no need in applying the Gaussian point spread function because the finite optical resolution is already accounted by integration over the size of the focal spot. Finally, the anticipated DOCP of the PL for both experiments that is shown in \autoref{fig:excitation&emission} was computed as 
    \begin{equation}\label{eq:DOCP}
        \mathrm{DOCP}=\frac{\tilde{\mathcal{I}}_{\sigma^{+}}-\tilde{\mathcal{I}}_{\sigma^{-}}}{\tilde{\mathcal{I}}_{\sigma^{+}}+\tilde{\mathcal{I}}_{\sigma^{-}}} \times \langle\mathrm{DOCP}^{\mathrm{bare}}_{\mathrm{PL}}\rangle
    \end{equation}
    where $\langle\mathrm{DOCP}^{\mathrm{bare}}_{\mathrm{PL}}\rangle$ was the averaged experimentally observed  DOCP of PL from embedded \ML{} without a GNP.
\section{Data availability}
    The data that support the plots within this paper and other findings of this study are available from the corresponding authors upon reasonable request.

\begin{acknowledgement}
    This work was funded by the Deutsche Forschungsgemeinschaft (DFG, German Research Foundataion) – CRC/SFB 1375 NOA ”Nonlinear Optics down to Atomic scales” (Project number 398816777), EXC 2051 (Project number 390713860), International Research Training Group 2675 “META-Active” (project number 437527638), and the Emmy Noether Program (Project number STA 1426/2-1). F.E. acknowledges the Bundesministerium für Bildung und Forschung (BMBF, Federal Ministry of Education and Research) support via NanoScopeFutur-2D (Project Number 13XP5053A). M.J.W. acknowledges support by the Australian Research Council (ARC) Centre of Excellence Grant No. CE170100039 and Schmidt Science Fellows, in partnership with Rhodes Trust. We thank M. Younesi and M. Rikers for providing the scanning electron micrographs and J. Gour and R. Schlegel for the preparation and coating of the substrates. We also thank U. Peschel, C. Rockstuhl and I. Fernandez-Corbaton for fruitful discussions on the symmetry properties of chiral emitters. Further, we thank A. Knorr, R. Salzwedel and L. Greten for fruitful discussions on dipole-dipole interactions in exciton-plasmon based systems.
\end{acknowledgement}

\section{Author contributions}
    T.B., Z.F., R.M., and I.S. conceived the research idea and designed the experiments; T.B., Z.F., and M.A. prepared the samples with the help of E.N., Z.G., H.K., A.G.; E.N., Z.G., A.G. performed the crystal growth under the supervision of A.T.; H.K. performed the thin-film integration under the supervision of F.E.; T.B. Z.F., and M.A. performed the experiments under the supervision of T.P. and I.S.; Z.F. and M.A. performed the numerical simulations; T.B. and Z.F. performed the data analysis and developed the physical model with the help of M.J.W.; T.B. and Z.F. prepared the figures and wrote the manuscript with the help of all authors; I.S. supervised this project.


\bibliography{references}

\end{document}


%
    %
    \noindent
    %

\section{Influence of the optical components on the detected polarization state of the PL}

\begin{figure}[h]
    \centering
    \includegraphics[width=\linewidth]{"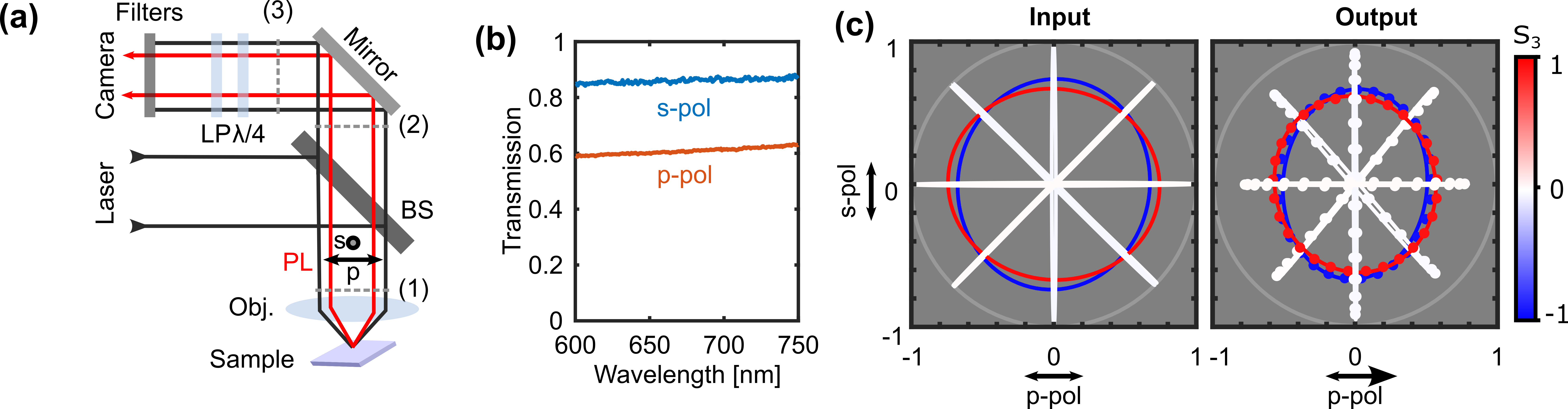"}
    \caption{\textbf{Stokes polarimetry in the cryostat setup.} (a) Sketch of the optical setup in reflection mode. The numbers in round brackets next to the gray dashed lines denote different positions. (b) Intensity transmission of s- and p-polarized light over a broad spectral range. (c) Polarization ellipses of the 6 analyzed cases averaged over the wavelength range from 650 nm up to 670 nm. Here, "Input" refers to the polarization state measured without a BS while "Output" refers to the polarization state after passing the BS. The circles in the "Output" plot show the theoretically calculated ellipses based on the estimated Müller matrix and the input state. The line color encodes the corresponding DOCP for clear differentiation between left- and right-handed circularly polarized light. 
    }{\phantomsubcaption\label{fig:SI_Beam_Splitter_a}}
    {\phantomsubcaption\label{fig:SI_Beam_Splitter_b}}
    {\phantomsubcaption\label{fig:SI_Beam_Splitter_c}}
    {\phantomsubcaption\label{fig:SI_Beam_Splitter_d}}
\end{figure}

Cryogenic photoluminescence (PL) microscopy experiments were performed in reflection mode, as depicted in Figure~\ref{fig:SI_Beam_Splitter_a}.
In this setup, the incoming laser beam is first deflected by a non-polarizing 30:70 beam splitter (BS) and then directed onto the sample inside the cryostat via an objective lens. The polarization state of the laser is routinely evaluated immediately after the BS, at a point marked as position 1 in Figure~\ref{fig:SI_Beam_Splitter_a}. The PL emitted from the sample is recollected by the same objective, transmitted through the BS, and then redirected by a mirror. Subsequently, the PL is analysed by a quarter-wave plate (QWP) and a linear polarizer (LP). This arrangement was motivated by the specific design of the cryostat chamber and the restrictions of the laboratory space. We will further investigate the influence of the BS and the mirror on the measured degree of circular polarization (DOCP) of the PL in such a setup.

\subsection{Beam splitter}
    First, we have analyzed the BS in a custom-build white-light spectroscopy setup for near-zeroth order transmittance (NA$\approx 0.044$) and under 45° incidence angle. In this setup, the incoming white light is prepared in an arbitrary polarization state, transmitted through the BS, and fiber-coupled into a spectrometer. By adding a rotatable QWP and fixed LP before the spectrometer fiber, we are able to measure the components of the Stokes vector (Fourier method) of the transmitted light. 
    Initially, we characterized the transmitted light for six degenerate input polarization states, namely the four linear polarized states vertical (s-polarization), horizontal (p-polarization), diagonal and anti-diagonal and for left- and right-handed circular polarization. The respective polarization ellipses measured for the "Input" (without BS) and "Output" (with BS) are shown in Figure~\ref{fig:SI_Beam_Splitter_c}, respectively. By comparison, we mainly observe three aspects: (1) the polarization ellipses are squeezed with respect to the horizontal axis (p-polarization), (2) the ellipticity of linearly polarized input states does not change and (3) the polarization ellipses for circularly polarized input states are not rotated. To describe the polarization behaviour of the BS we employ the Müller matrix formalism, where the output Stokes vector is related to the input Stokes vector via  $\textbf{S}^{\mathrm{out}}=M_{BS}\textbf{S}^{in}$. Based on the aforementioned observations, we conclude that our BS acts as a linear polarizing element with anisotropic amplitude attenuation coefficients $p_s$ and $p_p$.  Thus, the corresponding Müller matrix should have the form~\cite{collett2005field}
    \begin{equation}
        M_{BS}  =\frac{1}{2}\begin{bmatrix}
            p_s^2+p_p^2 & p_s^2-p_p^2 & 0 & 0\\
            p_s^2 -p_p^2 & p_s^2+p_p^2 & 0 & 0\\
            0 & 0 & 2p_s p_p & 0\\
            0 & 0 & 0 & 2p_s p_p\\
        \end{bmatrix}.
    \end{equation}
    Here,  $p_s^2$ and $p_p^2$ are intensity transmittances, $p_s^2=T_s$ and $p_p^2=T_s$, which are shown in Figure~\ref{fig:SI_Beam_Splitter_b} for a wide spectral range from 600 nm to 750 nm. The transmission characteristics of the BS exhibited minimal dependence on wavelength, with average transmission coefficients for the intensity of  $T_s=$0.86 and $T_p=$0.61 for s- and p-polarized light, respectively. Based on these values, the corresponding Müller matrix is calculated in a straightforward manner.
    In Figure~\ref{fig:SI_Beam_Splitter_c} we observe a strong agreement between the measured output polarization ellipses (solid lines) and the transformed input ellipses (circles) using computed $M_{BS}$. To understand how the BS influences the detected DOCP of the PL, let us consider a partially circularly polarized light as an input, i.e. $\textbf{S}^{\mathrm{in}}=\left[1,0,0,\alpha\right]^\mathrm{T}$ where $\alpha\in(-1,1)$ is a degree of circular polarization (DOCP). After the BS at the position 2 in Figure~\ref{fig:SI_Beam_Splitter_a} we obtain: $\textbf{S}^{\mathrm{out}}=M_{BS}\textbf{S}^{in}=0.735\cdot\left[1,0.170,0,0.985\cdot\alpha\right]^\mathrm{T}$.
    Thus, the BS induces the constant positive offset of $S_1$, $S_2$ remains zero, while the DOCP, represented by the $S_3$, is only negligibly decreased by a factor of 0.985 (when normalized by $S_0$).

\subsection{Mirror}
    After the BS, the PL is reflected from a protected silver mirror under 45$^{\circ}$. The Müller matrix for reflection from metals can be given in terms of the reflection coefficients $r_s$ and $r_p$~\cite{gramatikov2020mueller}
    \begin{equation}
        M_{\mathrm{Mirror}}  =\frac{1}{2}\begin{bmatrix}
            r_s^2+r_p^2 & r_s^2-r_p^2 & 0 & 0\\
            r_s^2 -r_p^2 & r_s^2+r_p^2 & 0 & 0\\
            0 & 0 & 2r_s r_p\cos{\gamma} & -2r_s r_p\sin{\gamma}\\
            0 & 0 & 2r_s r_p\sin{\gamma} & 2r_s r_p\cos{\gamma}
        \end{bmatrix},
    \end{equation}
    where $\gamma=\gamma_s-\gamma_p$ is the phase offset between s- and p- polarized field components. According to the manufacturer, the reflectance of the mirror at the PL wavelength and for 45$^{\circ}$ incidence angle are $r_s^2=R_s\approx0.97$ and $r_p^2=R_p\approx0.95$. Considering the normalized Stokes vector after the BS as $\textbf{S}^{in}=\left[1,0.17,0,0.985\cdot\alpha\right]^\mathrm{T}$, it will be transformed by the mirror as follows $\textbf{S}^{\mathrm{out}}=M_{\mathrm{Mirror}}\textbf{S}^{in}=0.962\cdot\left[1,0.18,-0.983\cdot\alpha\sin{\gamma},0.983\cdot\alpha\cos{\gamma}\right]^\mathrm{T}$. This analysis predicts that the detected polarization state of the PL will be influenced by both optical elements. Specifically, $S_1$ acquires a positive offset that is independent of the DOCP, $S_2$ shifts proportionally to the DOCP, and crucially, $S_3$ remains directly proportional to the DOCP. 

\subsection{Stokes polarimetry for the laser beam}\label{sec:LaserStokes}

    To validate our analysis we assessed the polarization state of the laser prepared in $\sigma_{+}$/$\sigma_{-}$ state at the positions numbered as 1 and 3 in Figure~\ref{fig:SI_Beam_Splitter_a}. The normalized Stokes components ${\textbf{S}} = \left(1, S_1, S_2, S_3\right)$ were defined as follows:
    \begin{align*}
        S_1 = \frac{\mathcal{I}_s - \mathcal{I}_p}{\mathcal{I}_s + \mathcal{I}_p},\hspace{0.4cm}S_2 = \frac{\mathcal{I}_d - \mathcal{I}_a}{\mathcal{I}_d + \mathcal{I}_a},\hspace{0.4cm}S_3 = \frac{\mathcal{I}_{\sigma^+} - \mathcal{I}_{\sigma^-}}{\mathcal{I}_{\sigma^+} + \mathcal{I}_{\sigma^-}},
    \end{align*} 
    where $\mathcal{I}$ denotes the light intensity, while the subscript indicates the detection polarization: linear (s) or (p), linear diagonal (d)/ antidiagonal (a), as well as two circular polarizations $\sigma_{+}$/$\sigma_{-}$. The obtained results are summarized in the table below:
    \begin{center}
        \begin{table}
            \begin{tabular}
                {c|a|a|b|b|}  
                \textbf{S} & $\sigma^+$, Position 1 & $\sigma^+$, Position 3 & $\sigma^-$, Position 1 &  $\sigma^-$, Position 3 \\
                \hline
                 S1 & $0.003\pm0.003$& $0.256\pm 0.006$ & $-0.008\pm 0.003$ & $0.173\pm0.004$  \\ 
                 S2 & $0.006\pm0.003$& $0.142\pm 0.005$ &$0.01\pm0.003$ & $-0.178\pm0.004$  \\ 
                 S3 & $0.977\pm0.003$ & $0.839\pm 0.003$ &$-0.972\pm0.005$ &  $-0.8673\pm0.002$ 
            \end{tabular}
        \end{table}
    \end{center}
    These measurements agree well with our expectations. Namely, the laser, initially prepared in an almost perfect $\sigma^+$ (or $\sigma^-$) polarization state at position 1, undergoes changes due to the BS and the mirror. This results in $S_1$ becoming positive, close to 0.18 as anticipated, $S_2$ shifting in proportion to the degree of circular polarization (DOCP), and $S_3$ showing a slight reduction. We attribute the minor discrepancies from the expected outcomes to misalignments of the BS and the mirror, as well as to unaccounted polarization effects from the objective lens.

\section{Polarization-resolved cryo-PL microscopy}

    We studied the modification of the farfield DOCP of valley-specific \textcolor{red}{\st{chiral}} emission from monolayer molybdenum disulphide (\ML{}) by resonant gold nanoparticles (GNP). In order to infer changes in the degree of polarization from the DOCP alone, linear polarization components must vanish identically. Hence, we have performed full Stokes-polarimetric measurements of the photoluminescence (PL) from \ML{} decorated with GNPs.
\begin{figure}[h]
		\centering
		\includegraphics[width=16cm]{"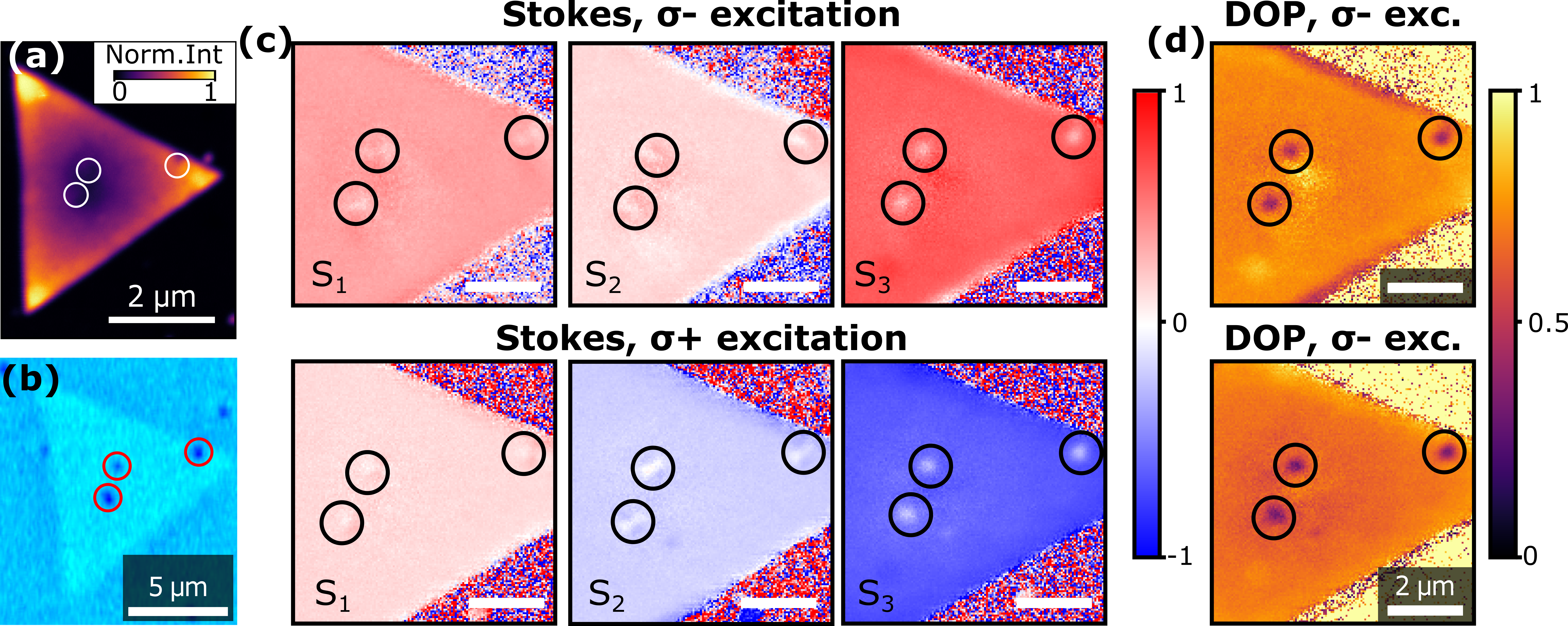"}
		{\phantomsubcaption\label{fig:SI_stokes_a}}
		{\phantomsubcaption\label{fig:SI_stokes_b}}
        {\phantomsubcaption\label{fig:SI_stokes_c}}
        {\phantomsubcaption\label{fig:SI_stokes_d}}
		\caption{
			\textbf{Polarization resolved photoluminescence imaging.} (a) Measured confocal scan of the total emission intensity $\mathcal{I}_{\text{tot}}$ from embedded \ML{} decorated with monodispersed GNPs upon $\sigma^+$ excitation and collected through a \SI{660}{\nano\metre} bandpass filter. (b) Optical microscopy image of the same sample. In both images the position of the GNPs is indicated by circles. (c) Measured confocal scans of the Stokes parameters of emission from the same sample upon $\sigma^-$ (top row) and $\sigma^+$ (bottom row) excitation. (d) Corresponding degree of polarization (DOP) for two excitation polarizations.  All scale bars in (c) and (d) represent 2\,µm.
		}\label{fig:SI_stokes}
	\end{figure}
   
    \autoref{fig:SI_stokes_a} shows the measured total emission intensity ${\mathcal{I}_\text{tot}}=\left(\mathcal{I}_{\sigma^+} + \mathcal{I}_{\sigma^-}\right)$ of a single crystal of \ML{} decorated by GNPs. \autoref{fig:SI_stokes_b} shows an optical microscopy image of the same sample and in both cases the GNPs are marked by circles, respectively. The GNPs have a negligible effect on the total emission intensity of \ML{} as no significant variation in brightness can be associated with the positions of the GNPs. In contrast, when calculating the respective Stokes parameter scans, as shown in \autoref{fig:SI_stokes_c} for $\sigma^-$ (top row) and $\sigma^+$ (bottom row) polarized excitation, we find a clear modulation of the polarization properties of emission from \ML{} mediated by the GNPs. For the DOCP or $S_3$ (right column), we find again a strong reduction as described in the main text. Furthermore, we observe the offsets of $S_1$ and $S_2$ induced by the utilized optical components as detailed in the previous section. These offsets align well with the behaviour of the initially circularly polarized laser light that passes through the same optical components (see subsection~\ref{sec:LaserStokes}). 
    At the center of the GNPs, $S_2$ approaches nearly zero, affected by the DOCP of the circularly polarized PL due to mixing of the Stokes vector components by the mirror. Around the nanoparticles, we observe subtle lobes of enhanced $S_1$ and $S_2$ indicating that it induces a mild linear polarization in its vicinity. Nonetheless, complete depolarization is evident at the center, as depicted in ~\autoref{fig:SI_stokes_d}, where the degree of polarization is calculated using the formula:
    \begin{equation*}
        \mathrm{DOP} \equiv\sqrt{S_1^2+S_2^2+S_3^3}.
    \end{equation*}
    This calculation reveals that the minor increase in linear polarization around the nanoparticle has a negligible impact on the overall degree of polarization, which is primarily affected by the decrease in $S_3$ Therefore, the alteration in $S_3$ is directly linked to changes in the degree of polarization.

\section{Polarization properties of valley-selective excitonic emitters}

For modelling emitters of circularly polarized light mainly two dipolar emitter types are discussed in literature:~\cite{fernandez2013role, zambrana2016tailoring, eismann2018exciting} (i) the rotating electric dipole $\Vec{\mathbf{p}}_{\sfrac{K}{K'}}=\Vec{\mathbf{p}}_x~\pm ~i\Vec{\mathbf{p}}_y$ where the spin-angular momentum of the emitter determines the sense of rotation and (ii) the chiral dipole $\Vec{\mathbf{p}}_{\sigma^{\pm}}=\Vec{\mathbf{p}}_x~\pm  
      ~i\Vec{\mathbf{m}}_x$ whose emission has a defined helicity due to duality invariance. In the context of 1L-TMDs, strong spin-orbit coupling as well as a broken inversion symmetry lead to valley-contrasting selection rules, i.e. the circularly-polarized external field $\sigma^{\pm}$ excites valley-polarized excitons, commonly approximated by rotating electric dipoles. 
\begin{figure}[h]
    \centering
    \includegraphics[width=13cm]{"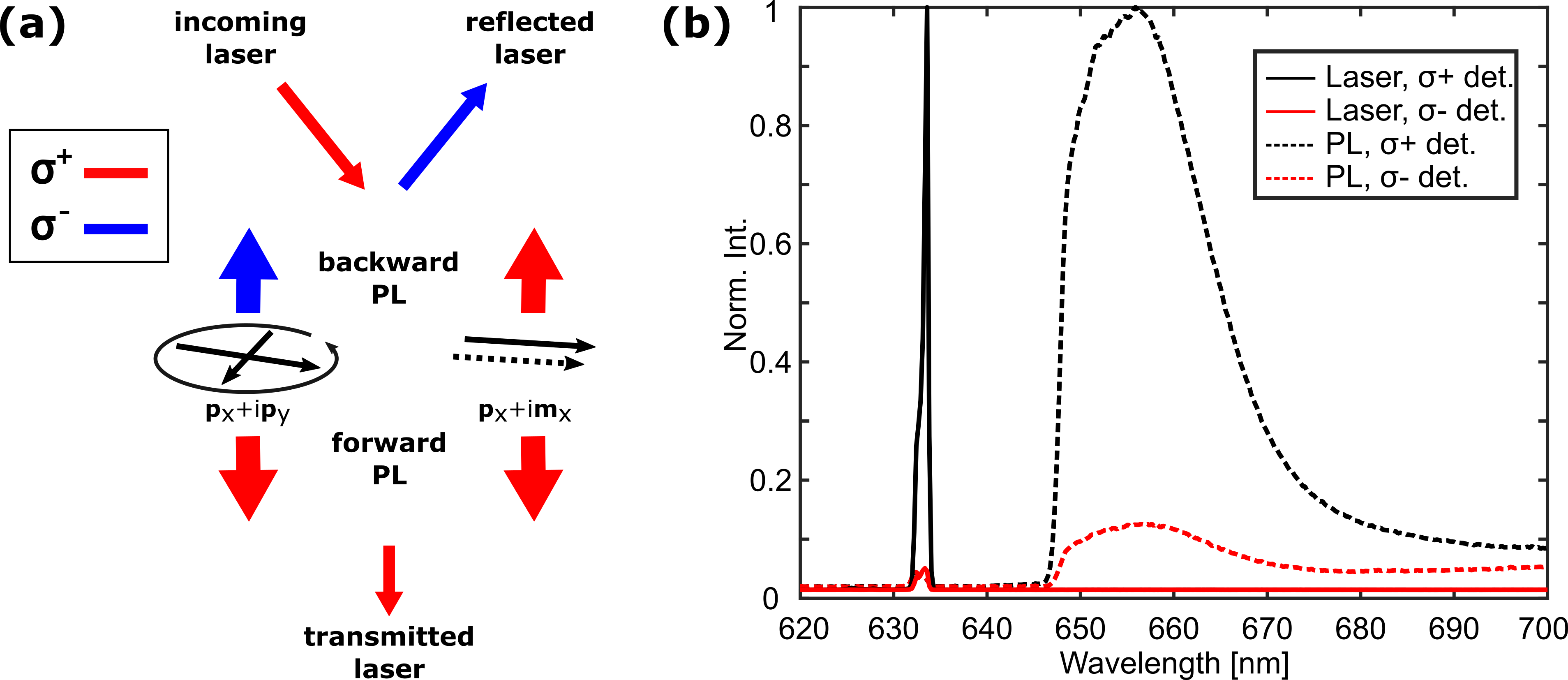"}
    {\phantomsubcaption\label{fig:SI_Spectra_PL_Laser_a}}
    {\phantomsubcaption\label{fig:SI_Spectra_PL_Laser_b}}
    \caption{
        \textbf{Comparison of the polarization contrast of reflected laser and emission.} (a) Sketch indicating the circular polarization of excitation light and emitted PL for the case of a rotating electric dipole (left side) and a chiral dipole (right side) upon $\sigma^+$ polarized excitation. (b) Normalized spectra of the $\sigma^+$-polarized HeNe laser detected after the reflection from the sample (solid lines) and of the photoluminescence from the bare \ML{} excited by this laser (dashed) for $\sigma_+$ (black) and $\sigma_-$ (red) detection polarizations. 
    }\label{fig:SI_Spectra_PL_Laser}
\end{figure}
The inset of Figure~5~(a) of the main text shows the farfield intensity distribution and respective DOCP of a single counterclockwise rotating electric dipole.
Such a rotating dipole emits circularly polarized light with opposite helicities into different half spaces with an average of exactly zero (as opposed to a chiral dipole whose helicity averages to $\pm 1$).\\
Here, we want to confirm experimentally that the polarization of the PL from a bare \ML{} qualitatively follows the behaviour of a rotating electric dipole. In reflection geometry and given a fixed excitation polarization, we can compare the polarization state of the valley-selective PL emitted in backwards direction to the that of the reflected laser beam as sketched in~\autoref{fig:SI_Spectra_PL_Laser_a} for a $\sigma^+$ polarized excitation. The rotating dipole (left side) inherits the spin-angular momentum from the excitation field and therefore emits light with the same handedness as the incoming laser beam into forward direction but with opposite handedness into backward direction. Note that upon the reflection from the sample, the circular polarization of the incoming laser also flips, i.e. $\sigma^+\longrightarrow \sigma^-$. Hence, the reflected laser light and the backwards emitted PL from a rotating dipole will be detected with the same circular polarization. Conversely, for the chiral dipole (right side) they would be detected with equal handedness as the emission polarization state does not depend on the emission direction.
\autoref{fig:SI_Spectra_PL_Laser_b} shows the measured polarization-resolved spectra of backwards emitted PL and reflected laser light from bare \ML{} upon excitation by a $\sigma^+$ polarized laser beam (HeNe, 633 nm). Both sets of spectra show a clear circular polarization contrast with each set of spectra favoring $\sigma^+$ polarization in detection. From this we can conclude that the polarization behaviour of bare \ML{} is qualitatively matching that of a rotating electric dipole. Note that the reflected laser light and the backwards emitted PL should appear with $\sigma^-$ polarization (as we have chosen $\sigma^+$ polarization for the excitation). However, before we detect it, the laser experiences an additional reflection from the mirror $\sigma^-\longrightarrow \sigma^+$ as shown in \autoref{fig:SI_Beam_Splitter_a}, position (3). As, a result, the excitation and detection polarization become the same.

\section{Dependence of PL depolarization on the spacer thickness}

\begin{figure}[h!]
		\centering
		\includegraphics[width=16cm]{"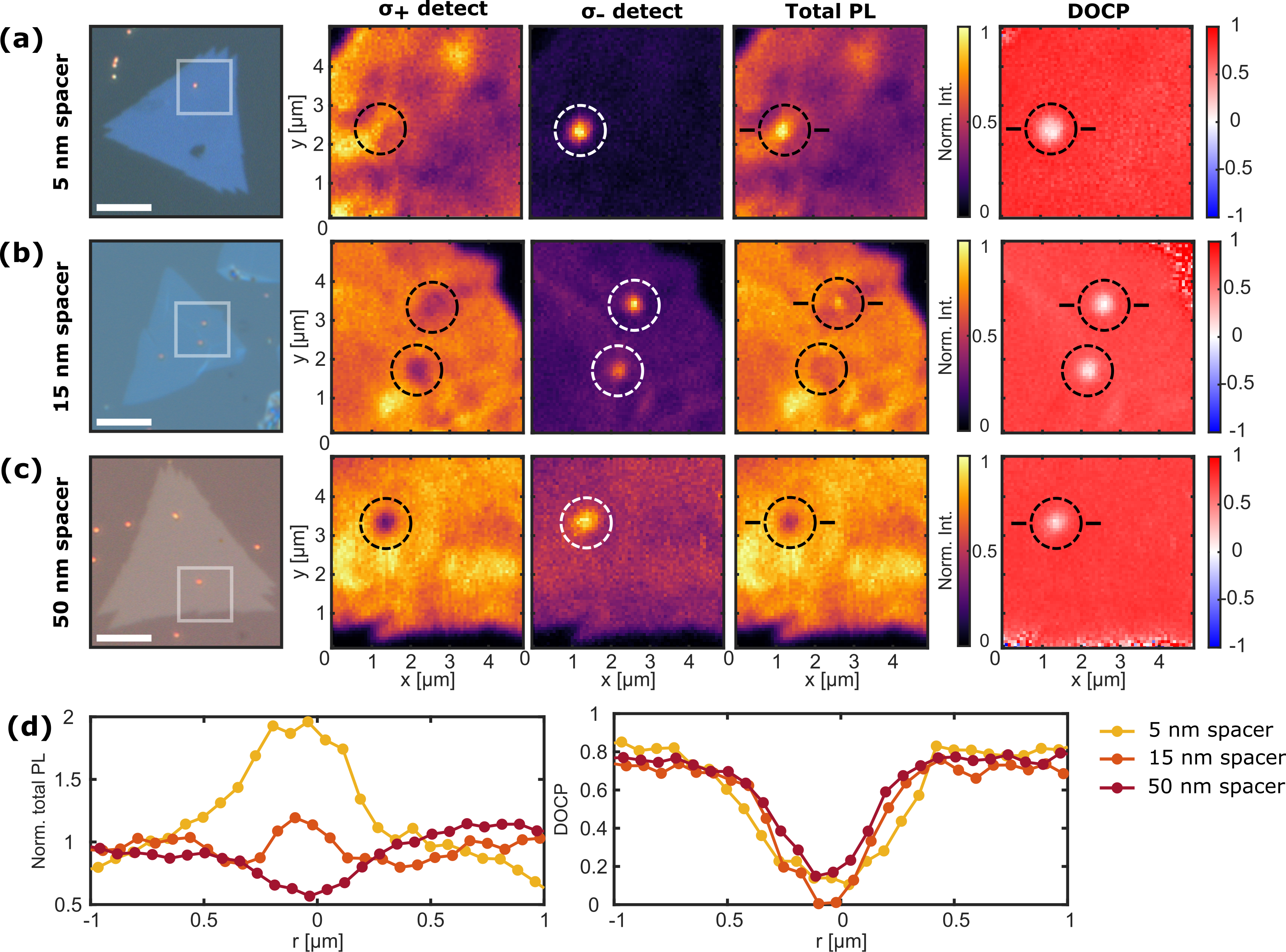"}
		\caption{
            \textbf{Influence of the spacer thickness.} (a-c) From left to right: Microscope images of the samples with highlighted scanning regions denoted by white squares, confocal scans showing the PL intensities for $\sigma^+$ and $\sigma^-$ excitation, the total PL intensity, and the DOCP obtained for varying spacer thicknesses (5 nm, 15 nm, and 50 nm). The positions of the GNPs are indicated with circles, and short horizontal lines adjacent to these circles mark the locations where the cross-sections depicted in (d) are taken. (d) Cross-sections through the center of the GNPs showing total PL intensity (left) and the DOCP for the samples illustrated in (a-c).
        }\label{fig:SI_Spacer_Thickness}
	\end{figure}

To explore the generality of a GNP's effect on the PL polarization we fabricated samples with different thicknesses of the SiO$_x$ spacer layer, namely \SI{5}{\nano\meter} and \SI{50}{\nano\meter} while maintaining identical conditions for other fabrication steps as described in the Methods section.  Figure~\ref{fig:SI_Spacer_Thickness}~(a-c) shows the result of confocal scanning microscopy of these two samples together with the case of \SI{15}{\nano\meter} spacer from the main text for comparison.  

We initially observe that the spacer thickness significantly affects the total PL intensity (see Figure~\ref{fig:SI_Spacer_Thickness}~(d), left). Specifically, the PL intensity is enhanced at the nanoparticle's position with a  \SI{5}{\nano\meter} spacer, remains nearly unchanged with a  \SI{15}{\nano\meter} spacer, and decreases with a  \SI{50}{\nano\meter} spacer. This variation in intensity can be explained by the interplay between GNP's absorption and nearfield enhancement, which counteract each other at shorter distances.  Despite variations in spacer thickness, all samples exhibit depolarization of the PL at the GNP location. Notably, the size of the depolarized (white) spot increases as the spacer thickness decreases (see Figure~\ref{fig:SI_Spacer_Thickness}~(d), right). This effect is attributed to the narrowing of the numerical aperture (NA) of the light cone interacting with the GNP as the distance from the source to the GNP increases.
\newpage

\section{Nearfield excitation intensity distribution for a GNP}

\begin{figure}[h]
		\centering
		\includegraphics[width=14cm]{"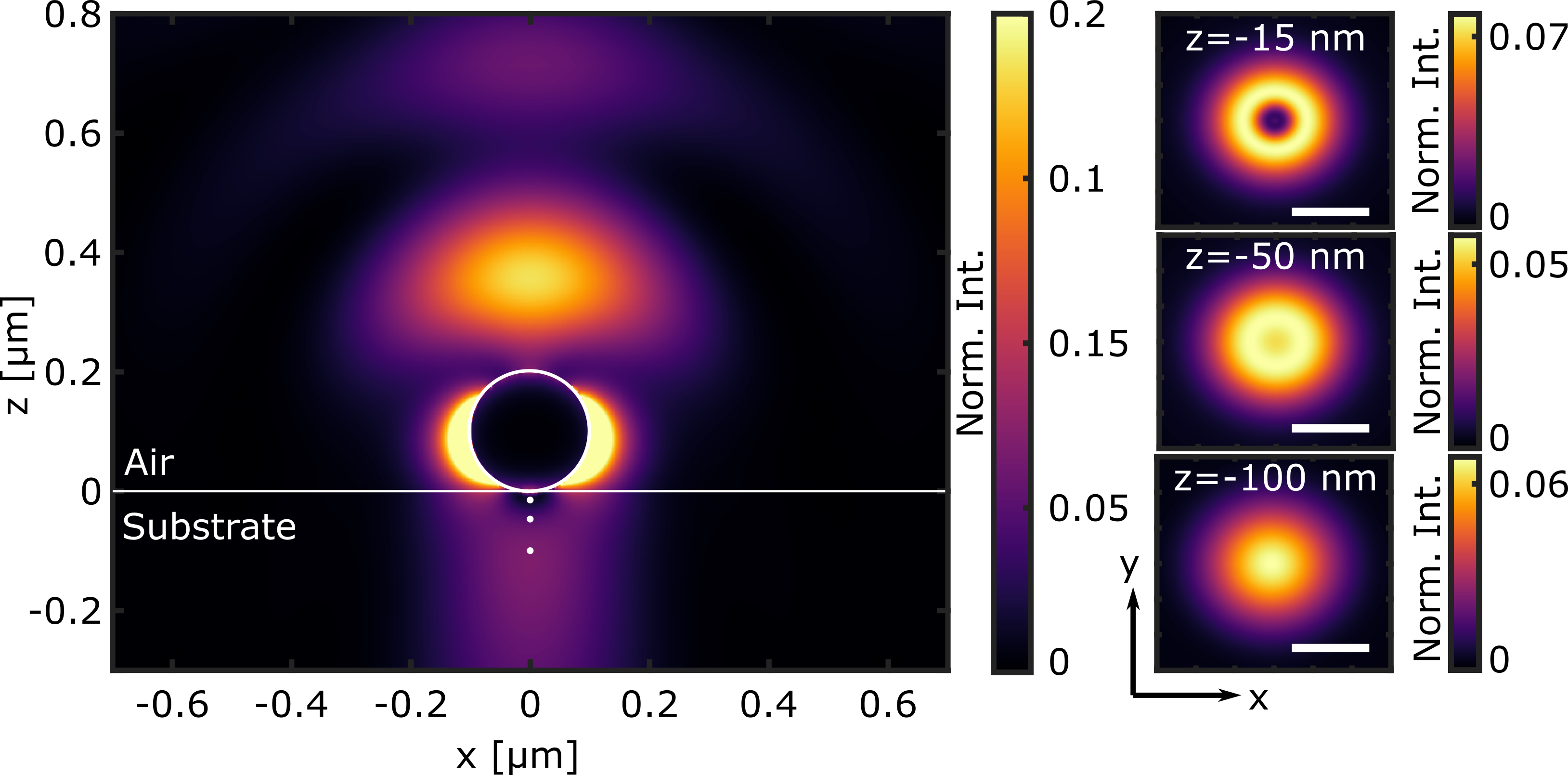"}
		\caption{
            \textbf{Nearfield intensity distribution of a gold nanoparticle on substrate.} Left: Numerically computed in-plane intensity $|E_x|^2+|E_y|^2$ of a GNP upon scattering of a tightly focused Gaussian beam shown in a x-z plane intersecting the center of a GNP. The white dots below the GNP correspond to $z=-15$~nm, $z=-50$~nm, and $z=-100$~nm. Right: in-plane intensity $|E_x|^2+|E_y|^2$ shown in x-y planes below the GNP at different z values. The scale bar corresponds to 200 nm.  
        }\label{fig:SI_Ring_Formation}
\end{figure}

Figure~\ref{fig:SI_Ring_Formation} illustrates the formation of the intensity pattern below a GNP upon scattering of the tightly focused Gaussian beam incident from $z=+\infty$. Here, we only considered the field components that are parallel to the substrate surface, i.e. $E_x$ and $E_y$.  It is evident that the discussed in the main text ring-like pattern below the GNP emerges at small distances away from the nanoparticle (<50 nm) while at larger distances the intensity distribution resembles that of a Gaussian beam. All simulation parameters are identical to those described in the main text.

\section{Emergence of cross-polarized nearfields below a GNP}
\begin{figure}[h]
		\centering
		\includegraphics[width=12cm]{"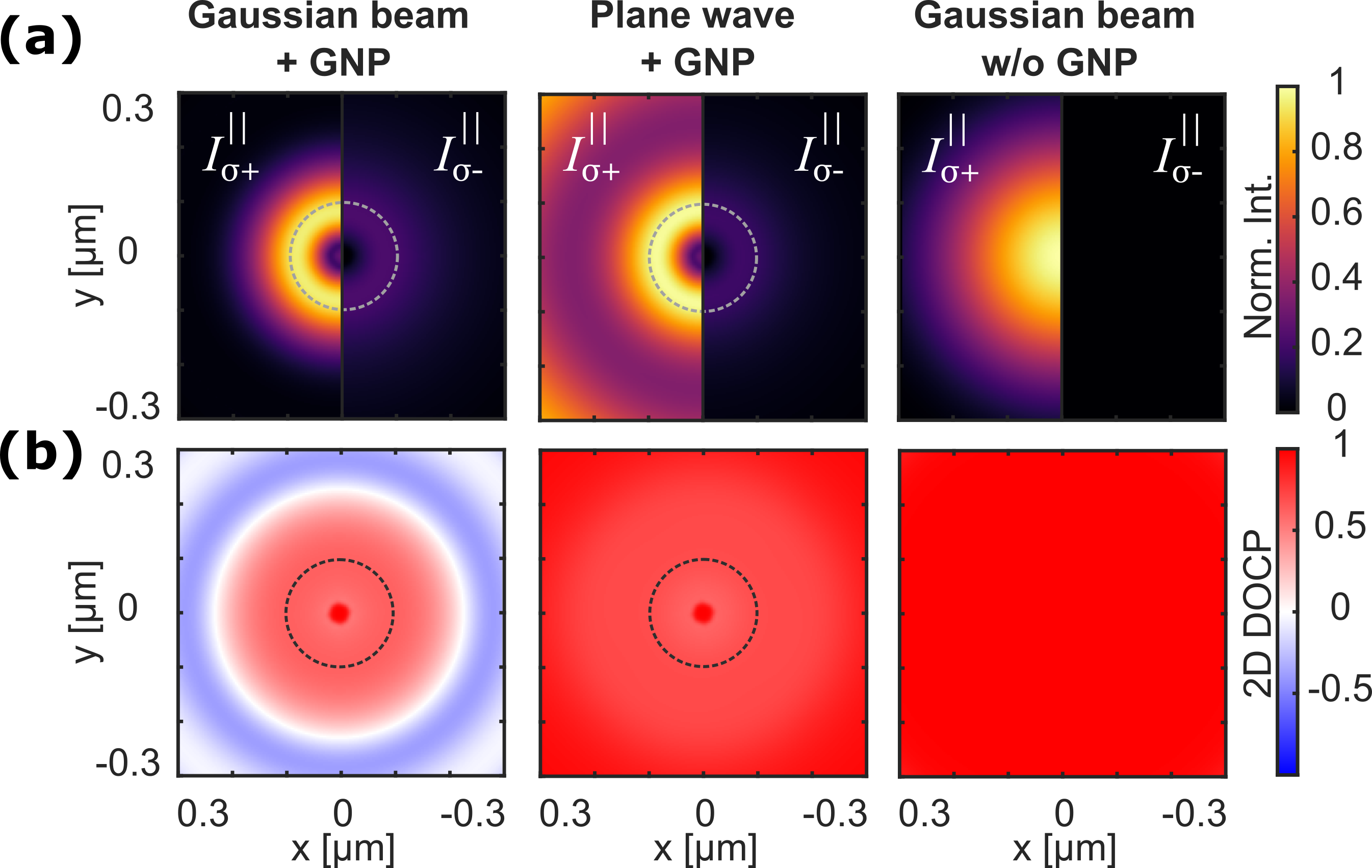"}
		\caption{
            \textbf{Helical intensity components in the nearfield of a gold nanoparticle.} (a) Numerically calculated $I^{||}_{\sigma^{+}}(x,y)$ and $I^{||}_{\sigma^{-}}(x,y)$ in the plane of the monolayer for different excitation scenarios. Only halves of the intensities are shown due to symmetry and both sides are normalized by the same value. The left column corresponds to a focused Gaussian beam on a GNP, the middle column - to the case of a plane wave incident on a GNP, and right column - to the focused Gaussian beam in the absence of a GNP. (b) Resultant 2D-DOCP in x-y plane.  The dotted circles highlight the projected GNP edge.
        }
        {\phantomsubcaption\label{fig:SI_Near_Fields_a}}
		{\phantomsubcaption\label{fig:SI_Near_Fields_b}}
\end{figure}

In Figure~\ref{fig:SI_Near_Fields_a} we plot the intensities of the $\sigma^+$ and $\sigma^-$ polarized in-plane field components in the x-y plane at $z=-15$~nm. We compare the cases of a focused Gaussian beam and a plane wave incident on a GNP, as well as the case of a focused Gaussian beam without a GNP as a reference. The excitation field is always $\sigma^+$ polarized while other simulation parameters are kept identical to those described in the main text. In Figure~\ref{fig:SI_Near_Fields_b} we analyze the resultant 2D-DOCP. Note that in the presence of a GNP and under Gaussian-beam excitation, a ring of reversed 2D-DOCP is observed. The location of this ring is easily deducible from Figure 4 in the main text, where the curves $I^{||}_{\sigma^{-}}(r)$ and $I^{||}_{\sigma^{+}}(r)$ cross. Conversely, for a plane wave, the curves do not intersect and the 2D-DOCP always stays positive with a slight reduction in the vicinity of the GNP. Reference data without a GNP confirms that the cross-polarized in-plane fields, leading to reduction or even reversal of 2D-DOCP, emerge only in the presence of a GNP.

\bibliography{referencesSI}